\documentclass[sigconf,nonacm]{aamas}
\usepackage{balance} % for balancing columns on the final page

\usepackage[utf8]{inputenc} % allow utf-8 input
\usepackage[T1]{fontenc}    % use 8-bit T1 fonts
\usepackage{hyperref}       % hyperlinks
\usepackage{url}            % simple URL typesetting
\usepackage{booktabs}       % professional-quality tables
\usepackage{amsfonts}       % blackboard math symbols
\usepackage{nicefrac}       % compact symbols for 1/2, etc.
\usepackage{microtype}      % microtypography
\usepackage{xcolor}         % colors
\usepackage{graphicx}
\usepackage{amsmath}
\usepackage{multirow}
\usepackage{adjustbox}
\usepackage{amsmath}
\usepackage{wrapfig}
\usepackage{rotating}
\usepackage{array}
\usepackage{tabularx}
\usepackage{makecell}
\usepackage{longtable}
\usepackage{pdflscape}

\title{Coordination Matters: Process-Aware Evaluation of Cooperative Multi-Agent Reinforcement Learning}

\author{Maria Ana Cardei}
\email{cbr8ru@virginia.edu}
\affiliation{%
  \institution{University of Virginia}
  \city{Charlottesville}
  \state{Virginia}
  \country{USA}
}

\author{Matthew Landers}
\affiliation{%
  \institution{University of Virginia}
  \city{Charlottesville}
  \state{Virginia}
  \country{USA}
}

\author{Afsaneh Doryab}
\affiliation{%
  \institution{University of Virginia}
  \city{Charlottesville}
  \state{Virginia}
  \country{USA}
}

\begin{abstract}
Cooperative multi-agent reinforcement learning (MARL) benchmarks commonly
emphasize aggregate outcomes such as return, success rate, or completion time.
While essential, these metrics can obscure how agents coordinate and fail to
distinguish policies that achieve similar outcomes through different
interaction processes. We propose a \emph{process-aware evaluation framework}
that complements aggregate performance with diagnostics of interference,
coordinated task completion, participation, and coordination efficiency.
We evaluate the framework in STAT, a controlled spatial task-allocation
testbed, and examine its broader applicability in two established cooperative
MARL benchmarks. Across diverse value-based and actor-critic methods, we find policies
with nearly identical return can exhibit substantial differences in
coordination behavior. Controlled scaling in STAT reveals distinct process
signatures for spatial, task, and agent scaling, while a reward-shaping
ablation shows that explicitly penalizing assignment conflicts can alter
coordination behavior with little change in task return. Results in the
established benchmarks further show that the framework remains informative
under different coordination mechanisms.
These findings support process-aware evaluation as a complement to
outcome-based benchmarking. The framework provides a common structure for
selecting and interpreting environment-specific diagnostics that characterize
not only what cooperative policies achieve, but how they coordinate to achieve
it.
\footnote{Code is available at
\href{https://github.com/mariacardei/coordination_aware_MARL}
{\texttt{https://github.com/mariacardei/coordination\_aware\_MARL}}.}
\end{abstract}


\newcommand{\BibTeX}{\rm B\kern-.05em{\sc i\kern-.025em b}\kern-.08em\TeX}

\begin{document}

%%% The following commands remove the headers in your paper. For final 
%%% papers, these will be inserted during the pagination process.

%\pagestyle{fancy}
%\fancyhead{}

%%% The next command prints the information defined in the preamble.

\maketitle
\pagestyle{plain}
\thispagestyle{plain}

%%%%%%%%%%%%%%%%%%%%%%%%%%%%%%%%%%%%%%%%%%%%%%%%%%%%%%%%%%%%%%%%%%%%%%%%

% \section{Introduction}

\section{Introduction}
\label{sec:intro}
Cooperative multi-agent reinforcement learning (MARL) involves multiple agents learning to act in a shared environment toward a common objective \cite{canese2021multiagent,Oroojlooy2023}. Success depends not only on individual capabilities, but also on how effectively agents coordinate. Agents must avoid redundant behavior, synchronize interdependent actions, divide work, and adapt to one another. Yet cooperative MARL methods are still commonly evaluated primarily through aggregate outcomes such as return, success rate, or completion time \cite{lowe2017,samvelyan2019smac,rashid2018qmix,mahajan2019maven,wang2021qplex}. This emphasis is also evident in evaluation practice, where a meta-analysis of cooperative MARL studies found that return, reward, and win rate dominate reported benchmark metrics, with many environments evaluated using only a single metric type \cite{gorsane2022standardised}.

% Aggregate outcomes measure task performance, but provide limited insight into the coordination that produced it.
% Policies with similar return may rely on substantially different interaction patterns, while changes in return may reflect coordination failures, execution inefficiencies, information limitations, or environment-specific bottlenecks. Outcome-equivalent policies therefore may not be coordination-equivalent. 
Aggregate outcomes measure task performance but provide limited insight into
the coordination processes that produce it. Policies with similar return can
exhibit substantially different patterns of interference, participation,
synchronization, and coordination efficiency. Similar outcomes therefore do
not necessarily imply similar coordination behavior.
Process-level diagnostics provide a more interpretable account of MARL performance by exposing the behaviors and failure modes underlying aggregate outcomes. Evaluating cooperative MARL thus requires examining not only \emph{what} a team achieves, but also \emph{how} agents coordinate to achieve it.

% \begin{figure}[t]
%     \centering
%     \includegraphics[width=0.7\linewidth]{Figures/updated_scatter.pdf}
%     \caption{Policies with similar return can exhibit substantially different coordination behavior. Process-level diagnostics reveal information that aggregate task outcomes alone can conceal.}
%     \label{fig:teaser}
% \end{figure}

We introduce a \emph{process-aware evaluation framework} that complements aggregate outcomes with environment-specific diagnostics of coordination behavior. 
% Instead of prescribing universal metrics, 
The framework organizes diagnostics around four general coordination dimensions: \emph{interference and redundancy}, \emph{coordinated task completion}, \emph{participation and division of labor}, and \emph{coordination efficiency}. These dimensions provide a common basis for identifying and interpreting process measures across environments with different coordination mechanisms.

The framework also distinguishes coordination behavior from its downstream consequences. Coordination opportunities produce agent behaviors and intermediate effects, such as redundant actions, successful joint task execution, inactivity, or delay, that ultimately influence throughput, duration, and return. Examining these intermediate stages helps separate coordination failures from other sources of performance variation that aggregate outcomes may conflate.

We instantiate the framework first in STAT, the Spatial Task Allocation Testbed, a configurable commitment-constrained environment designed to isolate assignment coordination.
% under controlled changes in agents, tasks, and spatial scale. 
We then evaluate broader applicability in Level-Based Foraging (LBF) \cite{christianos2020shared} and Robotic Warehouse (RWARE) \cite{papoudakis2021benchmarking}, which introduce distinct coordination demands including capability matching, coordinated loading, navigation, congestion, and partial observability. Across these settings, process diagnostics reveal behavioral differences and failure modes that aggregate return alone does not capture, providing evidence that process-aware evaluation is informative across distinct coordination mechanisms and benchmark settings.

Our contributions are as follows:
\begin{enumerate}
    \item We introduce a process-aware evaluation framework for cooperative MARL that complements aggregate outcomes with environment-specific diagnostics organized around four coordination dimensions: 1) interference and redundancy, 2) coordinated task completion, 3) participation and division of labor, and 4) coordination efficiency.

    % \item We instantiate the framework in STAT, a controlled commitment-constrained spatial task-allocation testbed that supports systematic variation in agent count, task count, and spatial extent while keeping the underlying coordination mechanism fixed.
    \item We establish a complementary evaluation design that combines controlled diagnostic depth with cross-environment validation. STAT isolates assignment coordination under systematic interventions, while LBF \cite{christianos2020shared} and RWARE \cite{papoudakis2021benchmarking} test whether the framework remains informative under different coordination mechanisms.

    % \item We instantiate and evaluate the framework across three complementary
    % environments. We introduce STAT, a controlled commitment-constrained spatial
    % task-allocation testbed designed to make coordination processes directly
    % observable.
    % % and to support systematic variation in agent count, task count,
    % % and spatial extent. 
    % We then evaluate the same process-aware perspective in
    % the established Level-Based Foraging \cite{christianos2020shared} and Robotic Warehouse \cite{papoudakis2021benchmarking} benchmarks, which
    % expose different coordination mechanisms. These settings provide controlled
    % diagnostic depth and cross-environment evidence for the broader applicability
    % of the framework.

    \item Across value-based and actor-critic MARL methods, we show that
    policies with similar aggregate return can exhibit substantially different
    coordination processes. Controlled scaling across configurations reveals distinct process
    signatures for spatial, task, and agent scaling, while a reward-shaping
    ablation shows that explicit coordination penalties can alter learned behavior with little change in task return.
    
    % \item We evaluate the framework in Level-Based Foraging and Robotic Warehouse, where environment-specific diagnostics capture distinct coordination mechanisms and reveal behavioral differences that aggregate outcomes alone do not characterize, providing cross-environment evidence for the broader applicability of process-aware evaluation.
\end{enumerate}

\section{Related Work}
\label{sec:RW}
\paragraph{Coordination and Process-Level Evaluation.}
Prior work has shown that aggregate performance alone does not fully
characterize multi-agent coordination. The broader multi-agent systems
literature has proposed coordination-specific measures for complex team
behavior \cite{maheswaran2008predictability}, while MARL research has examined
behavioral properties such as role diversity \cite{hu2022policy} and
agent-level coordination \cite{zhang2021coordination}. Related work in
human-AI and human-team collaboration similarly emphasizes interaction
processes such as communication, interdependence, control sharing, and
division of labor as important complements to task outcomes
\cite{biswas2026helping,strittmatter2026collaboration,Bethapuri2026}. Building on this foundation, we develop a process-aware evaluation framework for cooperative MARL that organizes environment-specific diagnostics around
common dimensions of coordination and uses them to complement aggregate
benchmark outcomes.

\paragraph{Spatial Task Allocation.}
Spatial task allocation is a longstanding problem in multi-agent systems and
robotics, where agents must service spatially distributed tasks under
assignment, movement, and execution constraints. Prior work studies scalable
planning \cite{claes2015effective}, spatial and temporal constraints such as
deadlines and sequential execution \cite{amador2014dynamic}, geographic
worker-task matching in spatial crowdsourcing
\cite{ye2021task,li2023spatial,zhao2024task,feng2025spatial}, and
reinforcement-learning approaches to multi-robot task allocation and joint
allocation-navigation \cite{agrawal2022rtaw,agrawal2022dcmrta}. This
literature primarily focuses on improving allocation or planning policies.
In contrast, we use spatial task allocation as a controlled setting for
making coordination processes directly observable and systematically varying
the pressures under which coordination occurs.

\paragraph{Cooperative MARL Benchmarks.}
Cooperative MARL benchmarks provide standardized testbeds across domains
including StarCraft micromanagement \cite{samvelyan2019smac},
particle-world coordination \cite{lowe2017}, Level-Based Foraging
\cite{christianos2020shared}, Overcooked-style collaboration
\cite{Carroll2019}, and robotic warehouse coordination
\cite{papoudakis2021benchmarking}. These environments expose challenges such
as partial observability, credit assignment, navigation, collaborative
planning, capability matching, and congestion. However, benchmark comparisons
are still commonly summarized through aggregate outcomes such as return, win
rate, or success rate
\cite{samvelyan2019smac,papoudakis2021benchmarking,gorsane2022standardised}.
Such metrics characterize what a policy achieves but provide limited insight
into how coordination succeeds or fails. We therefore complement aggregate
benchmark outcomes with environment-specific process diagnostics and evaluate
this perspective across settings with distinct coordination mechanisms.

% Table~\ref{tab:benchmark_comparison} in Appendix~\ref{app:stat_details}
% summarizes relevant benchmark characteristics.

\section{Preliminaries}
\label{sec:preliminaries}
\label{sec:preliminaries}

We model cooperative MARL as a decentralized partially observable Markov
decision process (Dec-POMDP) with $n$ agents
\cite{oliehoek2016concise}. The environment is defined by
\[
\mathcal{M}
=
\left\langle
\mathcal{S},
\{\mathcal{A}_i\}_{i=1}^{n},
P,
R,
\{\mathcal{O}_i\}_{i=1}^{n},
O,
\gamma
\right\rangle,
\]
where $\mathcal{S}$ is the global state space, $\mathcal{A}_i$ is the
action space of agent $i$, $P(s_{t+1}\mid s_t,\mathbf{a}_t)$ is the
transition function, $R(s_t,\mathbf{a}_t)$ is the shared team reward,
$\mathcal{O}_i$ is the observation space of agent $i$, $O$ is the
observation function, and $\gamma\in[0,1)$ is the discount factor.

At timestep $t$, each agent $i$ receives an observation
$o_t^i\in\mathcal{O}_i$ and selects an action
$a_t^i\in\mathcal{A}_i$ according to its policy $\pi_i$. The resulting
joint action is
\[
\mathbf{a}_t
=
(a_t^1,\ldots,a_t^n)
\in
\mathcal{A}_1\times\cdots\times\mathcal{A}_n.
\]
Under partial observability, $\pi_i$ may condition on the agent's local
action-observation history $\tau_t^i$, such that
$\pi_i(a_t^i\mid\tau_t^i)$. Under full observability, each agent can observe the global state.

The joint policy
$\boldsymbol{\pi}=(\pi_1,\ldots,\pi_n)$ maximizes expected discounted
team return,
\[
J(\boldsymbol{\pi})
=
\mathbb{E}_{\boldsymbol{\pi},P}
\left[
\sum_{t=0}^{T-1}
\gamma^t R(s_t,\mathbf{a}_t)
\right],
\qquad
\boldsymbol{\pi}^{*}
=
\arg\max_{\boldsymbol{\pi}}J(\boldsymbol{\pi}).
\]

% We evaluate methods spanning centralized training with decentralized execution
% and fully decentralized learning, allowing us to compare policies with
% different training structures under a common cooperative objective.
% \section{Preliminaries}

% \section{Process-Aware Evaluation of Coordination}
\section{Process-Aware Evaluation of Coordination}
\label{sec:process_eval}

Aggregate outcomes characterize \emph{what} a cooperative team achieves, but
provide limited information about \emph{how} agents interact to produce that
outcome. We introduce a process-aware evaluation framework that complements
task-level performance with environment-specific diagnostics of coordination
behavior and its intermediate consequences.

For a policy $\pi$, let $O(\pi)$ denote an aggregate outcome measure, such as
return, and let
\[
\mathbf{D}(\pi)
=
[D_1(\pi),\ldots,D_K(\pi)]
\]
denote process diagnostics computed from trajectories generated by $\pi$.
Whereas $O(\pi)$ summarizes downstream task performance, $\mathbf{D}(\pi)$
retains information about the interactions and behaviors through which that
performance is produced.

This distinction is useful whether outcomes are similar or different. Policies
with similar outcomes may coordinate in substantially different ways, while
changes in outcome may arise from different underlying coordination processes.
We characterize two policies $\pi_i$ and $\pi_j$ as \emph{outcome-similar}
when
\[
\Delta_O(\pi_i,\pi_j)
=
|O(\pi_i)-O(\pi_j)|
\leq \epsilon_O,
\]
yet their process diagnostics may still differ substantially. Conversely,
coordination behavior may change with little corresponding change in aggregate
outcome. Process and outcome evidence should therefore be interpreted jointly,
rather than treating either as a complete description of cooperative
performance.

\subsection{Coordination Dimensions}

Coordination manifests through different observable events across environments.
% Instead of prescribing a fixed set of metrics, 
We organize diagnostics around
four general coordination dimensions:

\begin{itemize}
    \item \textbf{Interference and redundancy.}
    Measures overlapping, conflicting, or obstructive behavior that wastes
    coordination opportunities, such as duplicate task selection or movement interference.

    \item \textbf{Coordinated task completion.}
    Measures whether interdependent agent behavior successfully produces
    progress on shared tasks, including successful joint actions and task
    completion.

    \item \textbf{Participation and division of labor.}
    Measures how useful work is distributed across agents, tasks, or
    interaction partners.

    \item \textbf{Coordination efficiency.}
    Measures the cost of converting coordinated behavior into useful progress,
    including wasted actions, inactivity, delays, and throughput.
\end{itemize}

These dimensions are not mutually exclusive, and a single diagnostic may
inform more than one dimension. Their purpose is to provide a common
interpretive structure for environment-specific measures, rather than a
universal scalar measure of coordination quality.

\subsection{Applying the Framework}

Applying the framework involves three steps.

\begin{enumerate}
    \item \textbf{Identify coordination opportunities.}
    Identify states or events in which agents' decisions are interdependent
    and coordinated behavior can affect task progress. Where possible,
    diagnostics should be normalized by relevant coordination opportunities
    or exposure so that differences in length, activity level, or
    decision frequency do not confound comparison.

    \item \textbf{Operationalize observable coordination processes.}
    Select diagnostics that capture the relevant coordination dimensions at
    those opportunities. The same dimension may require different observable
    measures across environments because the underlying coordination
    mechanisms differ.

    \item \textbf{Interpret process diagnostics jointly with task outcomes.}
    No individual diagnostic defines coordination quality in isolation. For
    example, low interference may indicate effective coordination, but may
    also reflect low task engagement or inactivity. Diagnostics should
    therefore be interpreted together with task completion and aggregate
    outcomes. Agreement strengthens behavioral conclusions, while disagreement
    exposes coordination patterns that outcome-only evaluation would conceal.
\end{enumerate}

Conceptually, process-aware evaluation traces the progression from
\emph{coordination opportunity} to \emph{behavior}, \emph{immediate
consequence}, and ultimately \emph{task outcome}. Aggregate metrics summarize
the downstream result of this process, while process diagnostics preserve
information about how that result was produced.

% \section{Coordination-Aware Evaluation Design}
% \label{sec:main_approach}
% \input{paper/main_approach}
% % \section{Methods}

% \section{Evaluation Environments and Process Diagnostics}
\section{Evaluation Environments and Process Diagnostics}
\label{sec:environments}

We instantiate the framework across three complementary environments.
We introduce STAT as a controlled testbed for isolating assignment
coordination, and use the established LBF \cite{christianos2020shared} and
RWARE \cite{papoudakis2021benchmarking} benchmarks to examine whether the
same coordination dimensions remain informative under distinct interaction
mechanisms. Table~\ref{tab:environment_overview} in Appendix~\ref{app:stat_details}
summarizes the environments and their distinct coordination mechanisms.

% \begin{figure}[t]
%     \centering
% \includegraphics[width=0.7\linewidth]{Figures/STAT.pdf}
%     \caption{Illustration of STAT. Agents start at a fixed origin and coordinate to complete spatially distributed tasks efficiently.
%     % At each timestep, an agent may select a task, move toward an assigned task, execute a task, or remain idle.
%     }
%     \label{fig:STAT_pic}
% \end{figure}

\subsection{STAT: Controlled Assignment Coordination}
\label{sec:stat}

The Spatial Task Allocation Testbed (STAT) is a configurable cooperative
multi-agent environment for controlled analysis of coordination.
% (Figure \ref{fig:STAT_pic}).
STAT generalizes a prior emergency-response task-allocation environment
\cite{CardeiCASE2024,cardei2025factorizeddeepqnetworkcooperative} into a
domain-agnostic setting in which agents coordinate over spatially
distributed tasks.
% Rather than reproducing the full complexity of a
% specific application, 
Low-level path planning and collision dynamics are abstracted away so that assignment-level interactions and their consequences are directly observable.

A team of agents selects tasks, travels to them, and executes them with the
goal of completing the full task set efficiently. STAT uses action masking and a finite-state commitment structure (Figure \ref{fig:FSM}) to isolate high-level task-allocation coordination. Once an assignment is retained, an agent remains committed to it through movement and task execution before becoming available for another assignment. Action masking restricts each agent to actions valid for its current mode and removes tasks that are already assigned or completed from the selectable set. As a result, meaningful coordination is concentrated at discrete assignment opportunities rather than continuously across all timesteps.

When more than one agent selects the same task at an assignment opportunity, STAT applies retrospective conflict resolution.
% If multiple agents select
% the same task at the same timestep, 
The agent closest to the selected task retains the assignment, while the others lose that assignment opportunity and are forced to \textit{idle} for that timestep. The retained task is then marked as assigned and removed from the selectable set in subsequent timesteps. This fixed rule resolves duplicate assignments after agents have made their choices without changing that they selected the same task. Conflicts therefore remain directly observable as incompatible assignment decisions and impose an immediate process-level consequence by
wasting assignment opportunities.
% without preventing eventual task completion.

Let $n_t$ denote the number of agents currently making assignment decisions
and $m_t$ the number of currently selectable tasks at assignment opportunity
$t$. Although the total numbers of agents and tasks are fixed within a STAT
configuration, $n_t$ and $m_t$ vary over the course of an episode as agents
become committed and tasks are assigned or completed. The corresponding
assignment-level joint action space is
\[
|\mathcal{A}_{\mathrm{assign}}(t)| = m_t^{\,n_t},
\]
yielding a state-dependent assignment problem whose coordination demands are
concentrated at sparse, high-impact decision points.
STAT uses a fixed task-level reward across configurations that encourages
timely task completion through a per-step cost and time-dependent
task-completion rewards. The complete reward specification and additional
environment formulation details are provided in
Appendix~\ref{app:stat_details}.

% Each STAT configuration fixes the number of agents, number of tasks, and
% spatial extent for the duration of an episode. Across configurations, these
% factors can be varied systematically while preserving the same underlying
% assignment mechanism and task rules. This controlled design allows us to
% isolate how changes in problem scale affect coordination processes and
% whether those changes are reflected in aggregate task performance.

\subsubsection{STAT Process Diagnostics}
\label{sec:stat_diagnostics}

\begin{figure}[t]
    \centering
    \includegraphics[width=\linewidth]{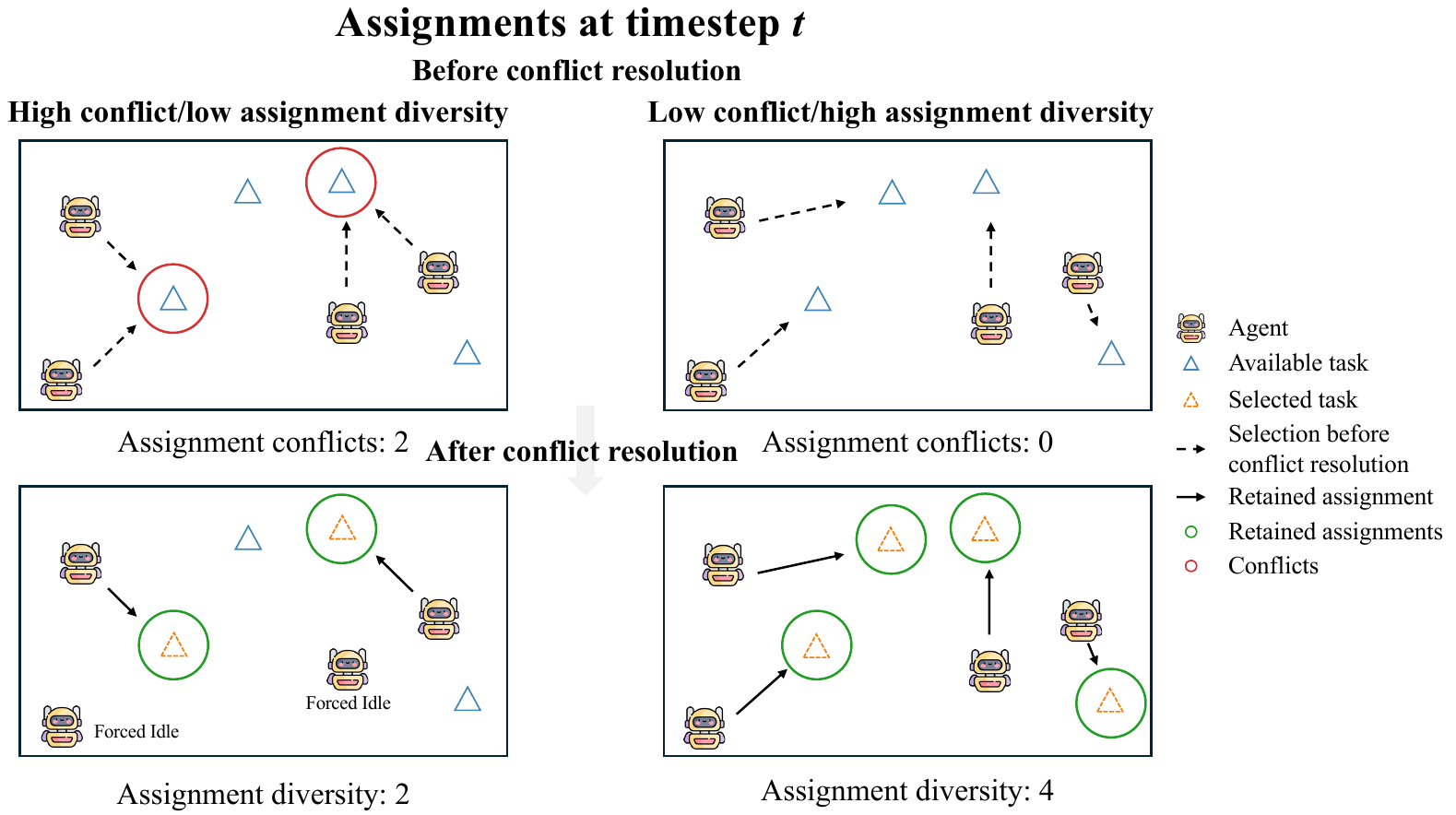}
    \caption{Illustration of STAT assignment diagnostics. The top row shows task
    selections before conflict resolution, and the bottom row shows the retained
    assignments after resolution. 
    % \textbf{Assignment conflicts} count tasks
    % selected by multiple agents before conflict resolution. \textbf{Assignment
    % diversity} measures the number of distinct retained task assignments after
    % resolution. Agents that lose conflict resolution become \textbf{forced
    % idle}, capturing an immediate efficiency consequence of incompatible
    % assignment decisions.
    }
    \label{fig:new_metrics}
\end{figure}

STAT makes assignment coordination directly observable, allowing process
diagnostics to be defined at the decision points where coordination occurs.
Let $\mathcal{T}_{\mathrm{opp}}$ denote the set of assignment opportunities
encountered during an episode. We instantiate the four coordination dimensions
introduced in Section~\ref{sec:process_eval} using measures of assignment
interference, task completion, division of labor, and coordination efficiency.
Figure~\ref{fig:new_metrics} illustrates the relationship between assignment
conflicts and retained assignment diversity.

\paragraph{Assignment Conflicts.}
To measure interference and redundant assignment, let $S_t$ denote the
multiset of tasks selected by decision-active agents at assignment opportunity
$t$. For each task $j$,
\[
n_t(j)
=
\sum_{s\in S_t}\mathbf{1}[s=j]
\]
denotes the number of agents selecting that task. The number of assignment
conflicts at opportunity $t$ is
\[
K_t
=
\sum_j \mathbf{1}[n_t(j)>1].
\]
Thus, $K_t$ counts the number of distinct tasks redundantly selected by
multiple agents before conflict resolution. Summing $K_t$ across assignment
opportunities yields the episode-level conflict count.

We additionally report normalized variants to support comparisons across
policies and configurations with different levels of coordination exposure.
Conflicts per decision opportunity normalize by
$|\mathcal{T}_{\mathrm{opp}}|$ and capture how frequently conflicts occur when
assignment decisions are made. Conflicts per task normalize by the total
number of tasks in the configuration and capture redundant assignment relative
to the size of the task set.

\paragraph{Forced Idle.}
Assignment conflicts impose an immediate consequence on agents that lose
conflict resolution. Let $F_t$ denote the number of agents forced to idle at
assignment opportunity $t$. Summing across assignment opportunities gives the
episode-level forced-idle count,
\[
F
=
\sum_{t\in\mathcal{T}_{\mathrm{opp}}} F_t.
\]

We report two normalized variants depending on the comparison of interest.
Forced idle per decision opportunity is
\[
F_{\mathrm{opp}}
=
\frac{1}{|\mathcal{T}_{\mathrm{opp}}|}
\sum_{t\in\mathcal{T}_{\mathrm{opp}}}F_t,
\]
which measures the average number of agents that lose an assignment
opportunity when coordination decisions occur. We also report the forced-idle
rate per timestep, which
% ,
% \[
% F_{\mathrm{step}}
% =
% \frac{F}{H},
% \]
% where $H$ is the episode length. 
measures conflict-induced inactivity
relative to total episode duration.

Because the \textit{idle} action is masked while selectable tasks remain,
forced idle reflects inactivity induced by conflict resolution rather than
voluntary non-participation. Assignment conflicts measure the occurrence of
redundant task selection, whereas forced idle measures its immediate severity
in lost agent participation.

\paragraph{Assignment Diversity.}
To characterize participation and division of labor, let
$A_t^{\mathrm{ret}}$ denote the retained task assignments after conflict
resolution. Assignment diversity is
\[
D_t
=
|A_t^{\mathrm{ret}}|,
\]
the number of distinct task assignments retained at opportunity $t$. Higher
values indicate that simultaneously decision-active agents distribute
themselves across a broader set of tasks. When comparing different team sizes,
we additionally normalize assignment diversity by the number of
decision-active agents.

\paragraph{Task Throughput.}
To measure how efficiently assignment decisions translate into task progress,
we report
\[
\rho_{\mathrm{STAT}}
=
\frac{M_{\mathrm{completed}}}{H},
\]
where $M_{\mathrm{completed}}$ is the number of tasks completed during an
episode of length $H$. Throughput captures downstream completion efficiency
and, when interpreted jointly with assignment-level diagnostics, helps
distinguish assignment failures from slower completion associated with travel
or commitment duration.

These diagnostics instantiate complementary dimensions of the
process-aware framework. Assignment conflicts capture interference and
redundancy, forced idle captures the immediate efficiency cost of incompatible
decisions, assignment diversity characterizes participation and division of
labor, and throughput captures coordinated task completion and downstream
efficiency. We interpret these diagnostics jointly with aggregate return.
% ; no
% individual metric is expected to characterize coordination quality in
% isolation.
% rather than treating any individual measure as a standalone measure
% of coordination quality.

\subsection{Established Benchmark Environments}
\label{sec:external_envs}

To complement STAT's controlled diagnostic setting, we evaluate the framework
in two established cooperative MARL benchmarks with distinct coordination
mechanisms.

\paragraph{Level-Based Foraging.}
LBF \cite{christianos2020shared} is a cooperative grid-world in which agents
collect food subject to capability constraints. Some food items require
multiple agents whose combined levels are sufficient, creating coordination
through capability matching and synchronized loading.

\paragraph{Robotic Warehouse.}
RWARE \cite{papoudakis2021benchmarking} is a partially observable warehouse
environment in which agents navigate shared aisles and transport requested
shelves to delivery locations. Coordination arises through shared-space
navigation, movement interference, congestion, and delivery behavior.

Table~\ref{tab:process_mapping} summarizes the environment-specific diagnostics
used to instantiate each process-aware dimension. Exact metric definitions are
provided in Appendix~\ref{app:envs}.

\begin{table}[t]
\centering
\caption{Environment-specific instantiations of the process-aware
coordination dimensions. Aggregate return is reported separately as an
outcome measure.}
\label{tab:process_mapping}
\small
\renewcommand{\arraystretch}{1.18}
\setlength{\tabcolsep}{5pt}
\begin{tabular}{
>{\raggedright\arraybackslash}p{0.22\linewidth}
>{\raggedright\arraybackslash}p{0.22\linewidth}
>{\raggedright\arraybackslash}p{0.22\linewidth}
>{\raggedright\arraybackslash}p{0.22\linewidth}}
\toprule
\textbf{Dimension} & \textbf{STAT} & \textbf{LBF} & \textbf{RWARE} \\
\midrule
Interference and redundancy
& Assignment conflicts
& Failed load-rate; synchronization failures
& Canceled movements \\

Coordinated task completion
& Task throughput
& Foods collected; collection throughput
& Delivery events; delivery throughput \\

Participation and division of labor
& Assignment diversity
& Participation balance; partner diversity
& Delivery contribution CV \\

Coordination efficiency
& Forced idle
& Synchronization delay
& Steps per delivery; NO-OP rate \\
\bottomrule
\end{tabular}
\end{table}

\section{Experimental Evaluation}
\label{sec:experimental_setup}
\subsection{Experimental Setup}
\label{sec:experimental_setup}

\subsubsection{Algorithms.}
We evaluate methods spanning centralized value learning, value decomposition,
independent learning, and actor--critic approaches. STAT includes DQN
\cite{mnih2015DQN}, FDQN
\cite{cardei2025factorizeddeepqnetworkcooperative}, VDN
\cite{Sunehag2017}, QMIX \cite{rashid2018qmix}, QTRAN
\cite{son2019qtran}, IQL \cite{Tan1993}, MAPPO and IPPO
\cite{yu2022surprising}, and MAA2C
\cite{mnih2016asynchronous,papoudakis2021benchmarking}. LBF evaluates MAPPO,
IPPO, QMIX, and IQL, while RWARE evaluates MAPPO and IPPO. Method details and
hyperparameters are provided in
Appendices~\ref{app:methods_evaluated} and~\ref{app:experimental_details}.

\begin{table}[t]
\centering
\small
\caption{STAT configurations used for controlled scaling analysis.}
\label{tab:problem_scales}
\begin{tabular}{lcccc}
\toprule
\textbf{Scale}
& \textbf{Agents}
& \textbf{Tasks}
& \textbf{Size}
& \textbf{Train Steps} \\
\midrule
\multirow{4}{*}{Baseline}
& 3 & 6  & $5\times3$   & 2M \\
& 3 & 6  & $10\times6$  & 2M \\
& 3 & 12 & $10\times6$  & 2M \\
& 5 & 12 & $10\times6$  & 2M \\
\midrule
\multirow{5}{*}{Extreme}
& 5 & 25  & $25\times15$ & 20M \\
& 5 & 25  & $50\times30$ & 20M \\
& 5 & 50  & $50\times30$ & 20M \\
& 5 & 100 & $50\times30$ & 20M \\
& 9 & 25  & $50\times30$ & 20M \\
\bottomrule
\end{tabular}
\end{table}

\subsubsection{Configurations.}
For STAT, we evaluate nine configurations that independently vary spatial
extent, task count, or agent count while preserving the underlying assignment
and commitment mechanisms. Increasing spatial extent changes travel burden,
increasing task count expands assignment availability, and increasing agent
count increases simultaneous assignment pressure.
Table~\ref{tab:problem_scales} summarizes the controlled comparisons.

For LBF, we evaluate two- and three-agent settings under full and partial
observability. For RWARE, we evaluate
\texttt{rware-tiny-2ag-v2},
\texttt{rware-tiny-4ag-v2}, and
\texttt{rware-small-4ag-v2}.
Exact environment identifiers, episode limits, reward settings, and additional
implementation details are provided in
Appendices~\ref{app:stat_details} and~\ref{app:envs}.

\subsubsection{Training and Evaluation.}
All experiments use five random seeds. During evaluation, learning is disabled
and policies use their deterministic evaluation rule. Unless otherwise stated,
we report means with 95\% confidence intervals. STAT uses 2M training steps
for Baseline configurations and 20M for Extreme configurations; RWARE uses
40M steps, while LBF training budgets vary by setting. Full training budgets,
evaluation frequencies, reward scalarization settings, and hyperparameters are
reported in Appendix~\ref{app:experimental_details}.

\subsection{Similar Outcomes Conceal Different Coordination Processes}
\label{sec:rq1}

We first test whether policies with similar aggregate performance exhibit
different coordination behavior. For each STAT configuration, we compare all
unordered pairs of evaluated methods, yielding 276
method-pair-configuration comparisons. Two policies are considered
\emph{outcome-similar} when their mean returns differ by at most 1\%, 2\%, or
3\%. For each such pair, process differences are measured as the ratio between
the larger and smaller mean diagnostic values.

{\setlength{\textfloatsep}{8pt}
\begin{table}[t]
\centering
\footnotesize
\setlength{\tabcolsep}{3pt}
\renewcommand{\arraystretch}{1.12}
\caption{Process differences among outcome-similar STAT policy pairs.
Counts indicate pairs with at least a $2\times$ gap in the corresponding
diagnostic.}
\begin{tabular}{cccc}
\toprule
\textbf{\begin{tabular}{c}Return\\Threshold\end{tabular}}
&
\textbf{\begin{tabular}{c}Outcome-\\Similar\end{tabular}}
&
\textbf{\begin{tabular}{c}$\geq 2\times$\\Conflicts\end{tabular}}
&
\textbf{\begin{tabular}{c}$\geq 2\times$\\Forced Idle\end{tabular}}
\\
\midrule
1\% & 121 & 31 & 38 \\
2\% & 199 & 58 & 80 \\
3\% & 244 & 76 & 103 \\
\bottomrule
\end{tabular}
\label{tab:rq1_thresholds}
\end{table} }

\begin{figure*}[t]
    \centering
    \includegraphics[width=\linewidth]{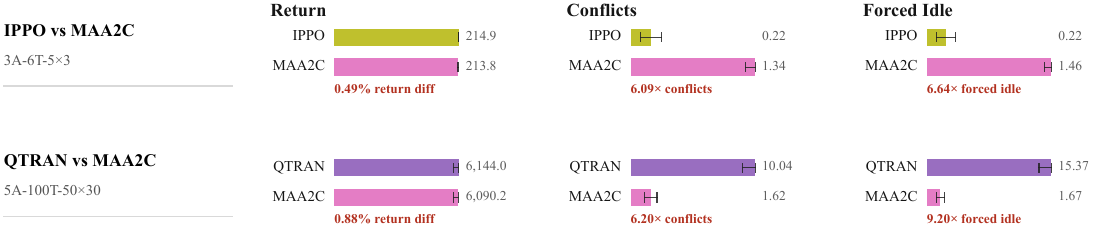}
    \caption{Example outcome-similar STAT policy pairs. Each row compares
    two methods whose mean returns differ by less than 1\%, while assignment
    conflicts and forced idle differ substantially. Bars show means across five
    seeds with 95\% confidence intervals.}
    \label{fig:rq1_examples}
\end{figure*}

Table~\ref{tab:rq1_thresholds} shows that substantial process differences are
common even under strict return-similarity criteria. At the 1\% threshold, 121
of 276 comparisons are outcome-similar. Among these, 38 (31.4\%) differ by at
least $2\times$ in forced idle and 31 (25.6\%) by at least $2\times$ in
assignment conflicts. More stringent process-gap thresholds show that 26 pairs
differ by at least $3\times$ in one or both diagnostics and 9 by at least
$5\times$ (Table~\ref{tab:rq1_gap_magnitude}). Similar patterns persist at the
2\% and 3\% return thresholds.
Figure~\ref{fig:rq1_examples} illustrates representative examples. Nearly identical returns can coexist
with several-fold differences in assignment conflicts and forced idle.
Although related, these diagnostics capture different aspects of the same
failure mode. Conflicts measure redundant task selections, while forced idle
measures the resulting loss of agent participation.

Overall, these results show that aggregate return can obscure variation in assignment interference and its immediate efficiency cost.
Assignment diversity and throughput vary less among outcome-similar pairs,
indicating that the strongest hidden differences in this analysis lie in
redundant assignment and conflict-induced inactivity.

\subsection{Process-Aware Metrics Can Characterize Scale Changes}
\label{sec:rq2}

\begin{figure*}[t]
    \centering
    \includegraphics[width=\textwidth]{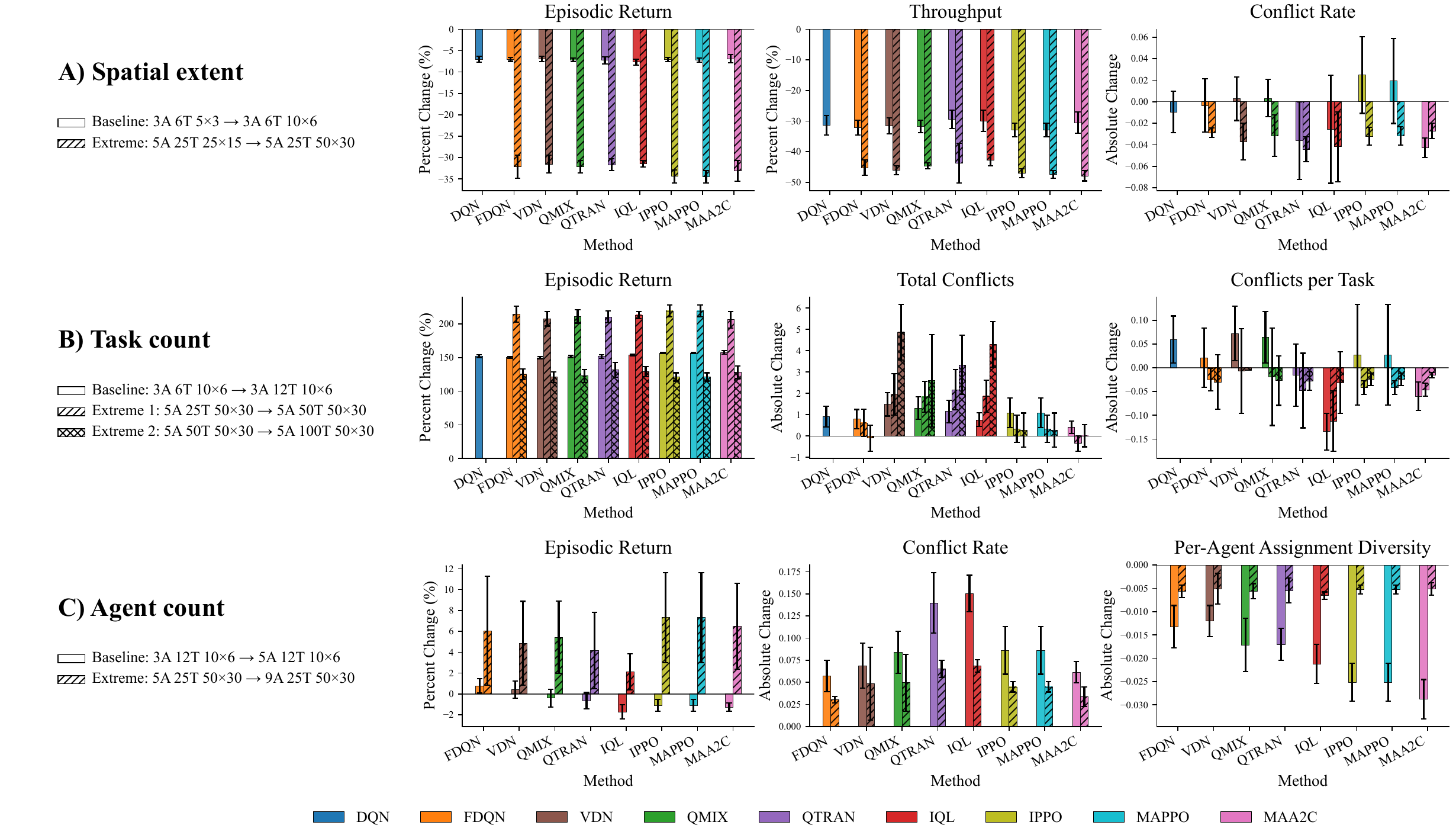}
    \caption{Process-aware analysis under controlled STAT scaling. Each row varies
    one configuration factor while holding the others fixed: (A) spatial extent,
    (B) task count, and (C) agent count. Bars show mean changes across five seeds
    with 95\% confidence intervals.}
    \label{fig:scaling_deltas}
\end{figure*}

We next examine whether process-aware diagnostics help characterize how
coordination changes under controlled scaling. Using paired STAT configurations from Table~\ref{tab:problem_scales}, we vary spatial extent, task
count, or agent count while holding the remaining configuration factors fixed.
Figure~\ref{fig:scaling_deltas} shows that these interventions produce distinct
process signatures that are not apparent from return alone.

% \paragraph{Increasing Spatial Extent.}
\textit{Increasing spatial extent }raises travel burden without changing team or task
size. As shown in Figure~\ref{fig:scaling_deltas}A, return and throughput
decrease substantially, while conflict-rate changes are comparatively small
and often negative. This pattern indicates that the primary effect of spatial
scaling is reduced completion efficiency and fewer assignment opportunities,
rather than increased redundant assignment.

% \paragraph{Increasing Task Count.}
\textit{Increasing task count} expands the assignment choice set and creates additional
opportunities for task completion. Figure~\ref{fig:scaling_deltas}B shows that
return generally increases, but total assignment conflicts also rise.
Conflicts per task change more modestly and can decrease, indicating that some
of the increase in absolute conflict is attributable to the larger task set.
Thus, improved aggregate performance can coexist with greater redundant
assignment.

% \paragraph{Increasing Agent Count.}
\textit{Increasing agent count }adds parallel capacity while also increasing the number
of simultaneous assignment decisions. Figure~\ref{fig:scaling_deltas}C shows
that the effect on return is method-dependent, whereas conflict pressure
generally increases and per-agent assignment diversity decreases. Additional
agents therefore do not automatically translate into proportionally broader
division of labor; they can also amplify redundant decision-making.

The three interventions show that scale does not induce a
single coordination failure mode. Spatial scaling primarily affects completion
efficiency, task scaling increases absolute assignment pressure, and agent
scaling introduces a tradeoff between parallel capacity and coordination
interference. Process-aware diagnostics therefore provide a more specific
account of scaling behavior than aggregate return alone.
Statistical tests for all controlled scaling comparisons, with Holm correction
within each comparison block, are reported in
Tables~\ref{tab:spatial_scaling_significance}-\ref{tab:agent_scaling_significance} in
Appendix~\ref{app:stat_results}.

\subsection{Explicit Coordination Penalties Change Behavior More Than Task Return}
\label{sec:rq3}

\begin{figure}[t]
    \centering
    \includegraphics[width=\linewidth]{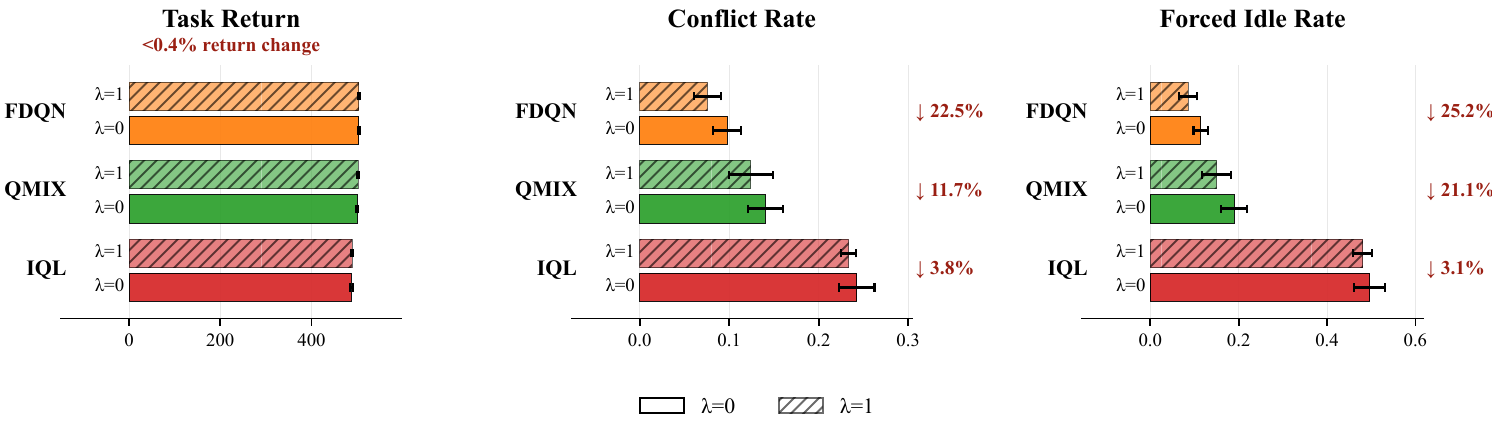}
    \caption{Reward-shaping ablation with an explicit conflict penalty in STAT.
    FDQN, QMIX, and IQL are trained with $\lambda=0$ or $\lambda=1$, where
    $\lambda$ penalizes agents that lose conflict resolution. Reported task
    return excludes the conflict penalty so that both conditions are compared
    under the same task-level objective. Bars show means across five seeds with
    95\% confidence intervals.}
    \label{fig:test_reward}
\end{figure}

A possible explanation for the process differences observed above is that
STAT's task reward does not explicitly penalize assignment conflicts. To test
this explanation, we conduct a reward-shaping ablation that adds an explicit
coordination penalty while holding the remaining training and evaluation
conditions fixed. In the 5-agent, 12-task, $10\times6$ STAT configuration, we
retrain FDQN, QMIX, and IQL using
\[
r_{\mathrm{train}}
=
r_{\mathrm{task}}
-
\lambda N_{\mathrm{loss}},
\]
where $N_{\mathrm{loss}}$ is the number of agents that lose conflict
resolution at a timestep. We compare the original objective
($\lambda=0$) with explicit conflict penalization ($\lambda=1$) using the
same training budget, seeds, layouts, and evaluation protocol.

% Because the shaped training rewards are not directly comparable, 
Both conditions are evaluated using the task return $r_{\mathrm{task}}$, with the
conflict penalty excluded. As shown in Figure~\ref{fig:test_reward}, explicitly
penalizing conflict-losing agents changes coordination behavior while leaving
task-level performance nearly unchanged. Conflict rate decreases by 22.4\%
for FDQN, 11.7\% for QMIX, and 3.8\% for IQL, with corresponding reductions
in forced-idle rate. In contrast, task return changes by less than 0.4\% for
every method.

The explicit coordination signal also does not eliminate differences across
methods. Under $\lambda=1$, conflict rates remain 0.0756 for FDQN, 0.1241 for
QMIX, and 0.2334 for IQL. Reward design therefore influences conflict
behavior, but does not fully explain the method-dependent coordination
patterns observed in STAT.

These results show that the outcome-process disconnect is not solely a
consequence of omitting an explicit coordination signal from the training
objective. Conflict shaping can reduce redundant assignment and
conflict-induced inactivity, yet these changes remain largely invisible to
task return. Process-aware diagnostics therefore reveal the effects of
coordination-focused interventions that return-only evaluation would largely
miss.

\subsection{Process-Aware Evaluation Remains Informative Across Environments}
\label{sec:rq4}

Finally, we test whether the process-aware framework remains informative under
coordination mechanisms that differ from STAT. We instantiate the dimensions
from Section~\ref{sec:process_eval} using environment-specific diagnostics in LBF
and RWARE. Full results across all evaluated configurations are reported in
Tables~\ref{tab:lbf_full_results} and~\ref{tab:rware_full_results} in
Appendix~\ref{app:lbf_rware}.

\paragraph{Level-Based Foraging.}
In the two-agent fully observable setting, MAPPO and IQL achieve similar return
(0.984 versus 0.994), yet IQL exhibits 65\% fewer failed loads and 16\% higher
throughput. 
Under partial observability, MAPPO, IPPO, and IQL all achieve returns between
0.981 and 0.990, while IQL has approximately half the failed-load rate,
substantially fewer synchronization failures, and about 20\% higher throughput
(Table~\ref{tab:lbf_full_results}). Thus, nearly identical task-level outcomes
can arise from substantially different coordination processes.
% The full results in Table~\ref{tab:lbf_full_results} show that this pattern is
% not limited to a single comparison. 
In the three-agent settings, participation
balance and partner diversity additionally characterize how successful
cooperation is distributed across the team. Under full observability, MAPPO
and IPPO achieve similarly strong outcomes and similar participation profiles,
whereas under partial observability, MAPPO achieves higher return, throughput,
participation balance, and partner diversity than IPPO. These results show that
process diagnostics remain informative both when aggregate outcomes are similar
and when they diverge.

\paragraph{Robotic Warehouse.}
RWARE provides a complementary case in which low interference alone does not
imply effective coordination. In \texttt{rware-small-4ag-v2}, MAPPO exhibits a
very low canceled-movement rate (0.0016) but produces only 0.120 delivery events
per episode and near-zero throughput (0.00024), whereas IPPO has higher movement
contention (0.0227) yet substantially more delivery events (2.708) and throughput
(0.00542). A similar pattern appears in \texttt{rware-tiny-4ag-v2}
(Table~\ref{tab:rware_full_results}). These results show that low contention may
reflect efficient coordination, but can also indicate limited task engagement,
and should therefore be interpreted jointly with task completion and inactivity.

\label{sec:discussion}
\section{Discussion}
\label{sec:discussion}

\subsection{Similar Outcomes Do Not Imply Similar Coordination}

Aggregate task performance provides an essential measure of whether a
cooperative policy succeeds, but it is a compressed representation of the
interaction process that produced that outcome. Our results show that policies
with similar outcomes may not be behaviorally equivalent. Different methods
can achieve comparable task performance while exhibiting substantially
different patterns of interference, synchronization, participation, and
coordination efficiency.
This distinction changes how close benchmark results should be interpreted.
When several methods achieve similar aggregate performance, this does not
necessarily imply that they have learned equivalent coordination strategies or
that meaningful behavioral differences have disappeared. Process-level
diagnostics can instead reveal tradeoffs that are hidden by a single outcome
measure. Algorithm comparisons should therefore consider both what policies
achieve and the interaction processes through which they achieve it.

\subsection{Process Diagnostics Require Joint and Contextual Interpretation}

The three environments provide complementary evidence for the framework.
STAT offers controlled diagnostic depth by isolating assignment coordination
and making conflicts and their immediate consequences directly observable,
while LBF and RWARE test the framework under distinct coordination mechanisms.
Accordingly, the same coordination dimensions manifest through different
observable processes: STAT exposes redundant assignment and conflict-induced
inactivity, LBF exposes failed or poorly synchronized joint actions, and RWARE
exposes movement interference, inactivity, and delivery efficiency. Across all
three environments, these diagnostics provide information beyond aggregate
performance by revealing how coordination succeeds, fails, or remains
underutilized.

This variation motivates the use of environment-specific diagnostics within a
common interpretive framework rather than a universal set of coordination
metrics. For the same reason, process-aware evaluation should not replace
return with a single alternative ``coordination score.'' Individual
diagnostics capture specific aspects of interaction and can be ambiguous when
considered in isolation. Low interference, for example, may reflect efficient
coordination, but the RWARE results show that it can also accompany limited
task engagement.
% Likewise, broad division of labor is informative only when
% distributed activity contributes to successful task completion.

Diagnostics may also capture different stages of the same coordination
process. In STAT, assignment conflicts measure the occurrence of redundant task
selection, while forced idle measures its immediate consequence in lost agent
participation. Although related, the two distinguish the frequency of a
coordination failure from its severity. Process diagnostics should therefore
be interpreted jointly with one another and with task outcomes. When comparing
settings with different scales or decision frequencies, they should also be
normalized by the relevant coordination opportunities.

\subsection{Separating Optimization from Evaluation}

A coordination behavior that appears undesirable during evaluation could be
encoded directly into the learning objective, but doing so does not eliminate
the need to measure it separately. Our conflict-penalty ablation shows that explicitly penalizing assignment conflicts changes learned coordination behavior while producing little change in the original task return, with method-dependent behavioral differences persisting across conditions.
Learning objectives and evaluation diagnostics therefore serve different
roles. The objective specifies what agents are optimized to achieve, whereas
diagnostics characterize how they behave while doing so. Requiring every
behavior of interest to appear in the reward risks conflating these roles and
may obscure tradeoffs among coordination processes.
This separation has implications for benchmark design. A process-aware
benchmark should identify the coordination opportunities created by the
environment, measure observable behaviors and consequences at those
opportunities, and retain aggregate task performance as a complementary
outcome measure. Controlled interventions, such as changing spatial extent,
task count, team size, or reward incentives, can then help determine which
parts of the coordination process are sensitive to particular environmental
or algorithmic changes.

\subsection{Limitations and Future Directions}

Our evaluation spans three environments with distinct coordination mechanisms,
but all are spatially grounded cooperative tasks. These settings provide
complementary evidence for the framework, but do not cover the full range of
cooperative interaction. Future work should examine communication-heavy
coordination, negotiation, distributed planning, heterogeneous roles and
capabilities, and collaborative decision-making without explicit spatial
structure.

The framework also does not automatically determine the appropriate
diagnostics for a new environment. Its dimensions provide a common interpretive
structure, but operationalization still requires identifying the relevant
coordination opportunities and observable interaction mechanisms. This
flexibility supports application across domains, but introduces researcher
judgment into metric selection. More systematic procedures for selecting and
validating diagnostics across benchmark variants remain an important direction.

Finally, most cross-method and cross-environment comparisons are observational
and should not be interpreted as causal explanations of performance
differences. Controlled STAT scaling and the reward-shaping ablation provide
targeted interventions on specific sources of coordination pressure, but do not
establish general causal relationships between individual process measures and
downstream outcomes. Establishing such relationships is an important direction
for future work.

% \section{Conclusion}
\label{sec:conclusion}
\section{Conclusion}
\label{sec:conclusion}

We introduced a process-aware framework for evaluating cooperative MARL that
complements aggregate task performance with diagnostics of interference,
coordinated task completion, participation, and coordination efficiency. Across STAT, Level-Based Foraging, and Robotic Warehouse, process diagnostics
reveal coordination differences that aggregate return can obscure, including
among policies with similar outcomes and under controlled changes in scale and
reward shaping. The same coordination dimensions remain informative
across distinct interaction mechanisms, while requiring environment-specific
operationalization and joint interpretation with task outcomes. The framework provides a 
structure for understanding not only what cooperative policies achieve, but
how agents coordinate to achieve it.

% \newpage

\begin{acks}
%NSF GRFP
This work was supported by the National Science Foundation Graduate Research Fellowship Program under grant
number 2234693. 
We used an icon from FlatIcon by author Freepik.

% If you wish to include any acknowledgments in your paper (e.g., to 
% people or funding agencies), please do so using the `\texttt{acks}' 
% environment. Note that the text of your acknowledgments will be omitted
% if you compile your document with the `\texttt{anonymous}' option.
\end{acks}

%%%%%%%%%%%%%%%%%%%%%%%%%%%%%%%%%%%%%%%%%%%%%%%%%%%%%%%%%%%%%%%%%%%%%%%%

%%% The next two lines define, first, the bibliography style to be 
%%% applied, and, second, the bibliography file to be used.

\bibliographystyle{ACM-Reference-Format} 
\bibliography{sample}

\newpage
\appendix

\section{Code Release}
\label{app:code_release}
% We release the STAT environment together with executable training and evaluation code for all methods considered in this paper. The release includes DQN and FDQN training/evaluation scripts, PyMARL integration for IQL, VDN, QMIX, QTRAN, and COMA, smoke-test configurations, and example commands for running methods on configurable STAT instances. This artifact is intended to support reproducibility, executable verification, and future extensions of STAT with additional methods or environment variants. Code is available at \href{https://anonymous.4open.science/r/coordination_aware_MARL-251F/}{\texttt{anonymous.4open.science/r/coordination\_aware\_MARL-251F}}.

We release the STAT environment together with executable training and
evaluation code for the methods and environment configurations used in this
study. The release includes DQN and FDQN implementations,
PyMARL/EPyMARL-based implementations of IQL, VDN, QMIX, QTRAN, MAPPO,
IPPO, and MAA2C \cite{samvelyan2019smac,papoudakis2021benchmarking},
configurable STAT instances, LBF and RWARE experiment configurations,
process-diagnostic logging, smoke-test settings, and example commands for
reproducing the main experiments. The artifact is intended to support
reproducibility, executable verification, and future extensions with
additional methods, diagnostics, or environment variants. Code is available at \\
\href{https://github.com/mariacardei/coordination_aware_MARL}
{\texttt{https://github.com/mariacardei/coordination\_aware\_MARL}}.
% \textcolor{red}{[TODO: add code to repo, add to anon repo and update link if needed]}

\section{STAT Environment Details}
\label{app:stat_details}
We provide additional details on STAT, the Spatial Task Allocation Testbed,
used to instantiate process-aware evaluation in commitment-constrained spatial
task allocation. STAT generalizes the victim-tagging environment used in prior
work \cite{cardei2025factorizeddeepqnetworkcooperative,CardeiCASE2024} into a
domain-general task-allocation testbed. Unlike the prior application-specific
formulation, STAT systematically varies agent count, task count, and spatial
extent while preserving the underlying assignment mechanism, allowing
coordination processes to be analyzed under controlled changes in problem
structure.

\subsection{Environment Parameters, State, and Observability}

A STAT instance is specified by
\[
\mathcal{E} = \langle n, m, W, H, K, v, \Theta_R \rangle,
\]
where $n$ is the number of agents, $m$ is the number of tasks, $W \times H$ defines the spatial grid, $K$ is the number of timesteps required to execute a task after arrival, $v$ is the agent movement speed, and $\Theta_R = \{R_0,\eta,\beta,\alpha,\lambda_{\mathrm{step}}\}$ denotes the reward parameters. The current benchmark is fully observable, and agents and tasks are homogeneous, but the same formulation can be extended to agent-specific speeds, task-specific execution times, heterogeneous task requirements, or partial observability in future work.

Let $\mathcal{N}=\{1,\ldots,n\}$ denote the set of agents and $\mathcal{M}=\{1,\ldots,m\}$ denote the set of tasks. All agents begin from a fixed origin, and tasks are spatially distributed across a 2D grid. At timestep $t$, agent $i \in \mathcal{N}$ has position $x_i(t) \in [0,W]\times[0,H]$, mode $q_i(t)$, and current assignment $g_i(t) \in \mathcal{M}\cup\{\varnothing\}$. Each task $j \in \mathcal{M}$ has a fixed location $y_j$ and status
\[
z_j(t)\in\{\textsc{available},\textsc{assigned},\textsc{completed}\}.
\]
The team objective is to complete all tasks efficiently. Task execution is abstracted as the agent remaining at the task location for a fixed number of timesteps. An episode terminates when all tasks are completed or when the environment reaches the maximum episode length. 

We use a fully observable setting so that comparisons across training and execution paradigms are not confounded by differences in information access. The global state includes agent--task distance information, each agent's current mode, and task status variables indicating whether each task is available, assigned, or completed. Although STAT is fully observable in this benchmark, its structure naturally supports partially observable variants in future work.

\subsection{Action Space and Finite-State Commitment}

\begin{figure}[h]
    \centering
\includegraphics[width=\linewidth]{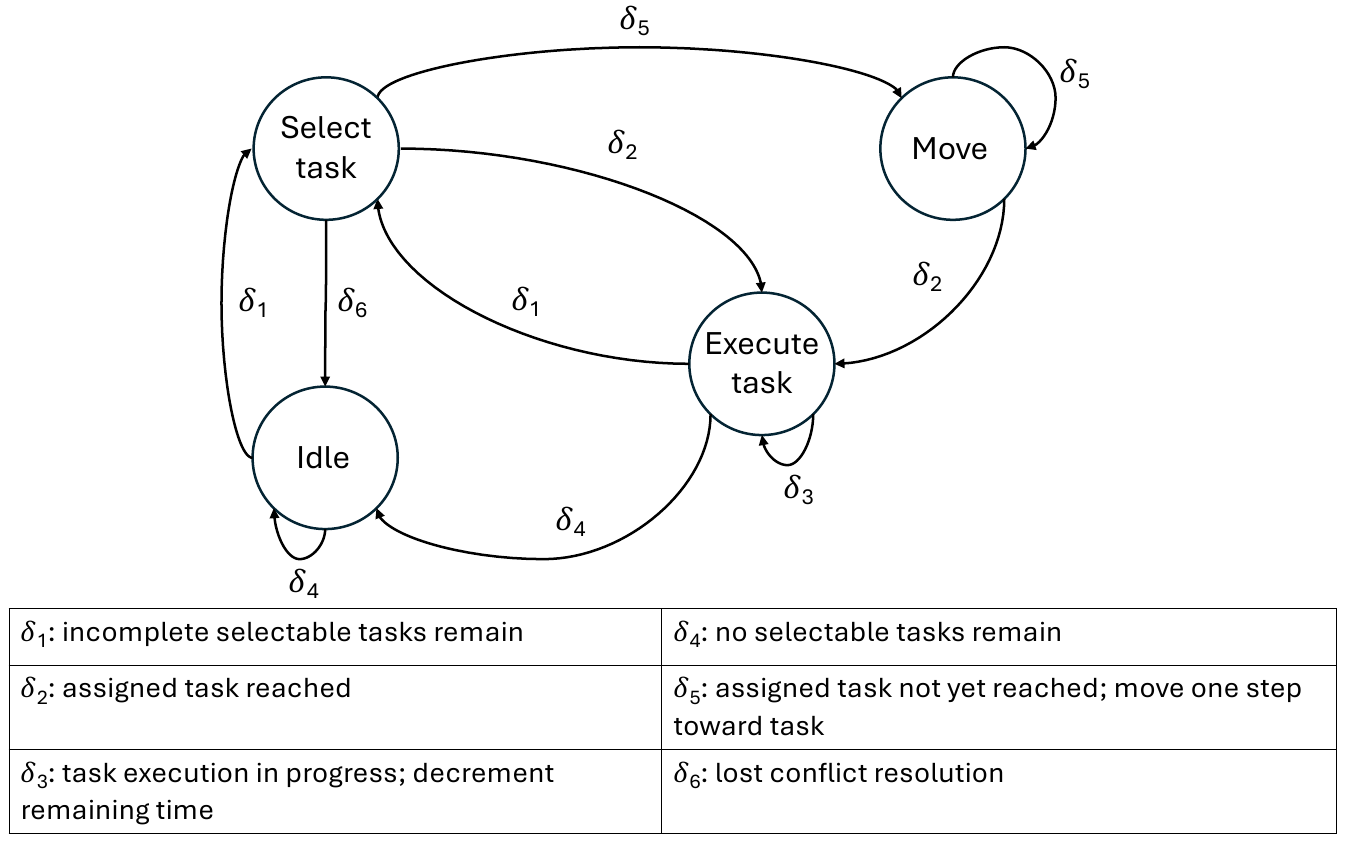}
    \caption{Finite-state commitment structure representing agent modes and valid transitions.}
    \label{fig:FSM}
\end{figure}

Each agent has a discrete action space consisting of \textsc{idle}, \textsc{move}, \textsc{execute}, and \textsc{select} actions. The \textsc{select} action expands into one action for each task:
\[
\mathcal{A}_i =
\{\textsc{idle},\textsc{move},\textsc{execute}\}
\cup
\{\textsc{select}(j): j\in\mathcal{M}\}.
\]
Thus, the nominal action set has size $3+m$. The valid action set is state-dependent and is enforced through action masking, as described in the next subsection.

Agents are governed by a finite-state commitment structure (Figure~\ref{fig:FSM}). We represent each agent's mode as
\[
q_i(t)\in \{\textsc{idle},\textsc{select task},\textsc{move},\textsc{execute task}\}.
\]
The \textsc{select task} mode is the decision mode in which an agent may choose among currently selectable tasks. Once a task is selected and retained after conflict resolution, the agent becomes committed to that task and transitions to \textsc{move}. The agent remains in \textsc{move} while it advances toward the assigned task. Once the assigned task is reached, the agent transitions to \textsc{execute task}, where it remains for a fixed number of timesteps \(K\) until the task is completed. After execution, the agent returns to \textsc{select task} if incomplete selectable tasks remain, or transitions to \textsc{idle} if no selectable tasks remain. If an agent loses conflict resolution after selecting a task, it transitions to \textsc{idle} for that timestep.

This design makes STAT a commitment-constrained task-allocation problem. The nominal assignment space is combinatorial, but the effective action space is state-dependent. As agents commit to tasks and as tasks become assigned or completed, the set of selectable tasks decreases. Therefore, the complexity of the assignment problem changes over the episode as early timesteps may contain many feasible task assignments, while later timesteps contain fewer meaningful choices. This property concentrates coordination pressure at sparse, high-impact decision points.
\subsection{Action Masking and State-Dependent Assignment Complexity}

STAT uses action masks to enforce the finite-state commitment structure and remove invalid actions. Let $\mathcal{V}_i(s_t)\subseteq \mathcal{A}_i$ denote the valid action set for agent $i$ in state $s_t$. If agent $i$ is in \textsc{select task} mode, then its valid task-selection actions are
\(
\mathcal{V}_i(s_t)=\{\textsc{select}(j): z_j(t)=\textsc{available}\}.
\)
If agent $i$ is in \textsc{move} mode and has not yet reached its assigned task, then the valid action set is restricted to
\(
\mathcal{V}_i(s_t)=\{\textsc{move}\}.
\)
If agent $i$ is in \textsc{execute task} mode, then the valid action set is restricted to
\(
\mathcal{V}_i(s_t)=\{\textsc{execute}\}.
\)
If no selectable tasks remain, the valid action set is restricted to
\(
\mathcal{V}_i(s_t)=\{\textsc{idle}\}.
\)
Completed or already assigned tasks are removed from the set of selectable task actions.

The masking removes invalid low-level choices so that the benchmark focuses on the high-level coordination problem of distributing agents across tasks. Without masking, a substantial part of the learning problem would involve discovering which actions are invalid in each state. With masking, the core challenge becomes whether agents make compatible assignment decisions when meaningful choices are available.

At an assignment decision point, let $m_t$ be the number of selectable tasks and $n_t$ be the number of agents currently in \textsc{select task} mode. The effective assignment-level joint action space is
\[
|\mathcal{A}_{\mathrm{assign}}(t)| = m_t^{n_t}.
\]
Thus, the assignment space grows combinatorially with the number of agents simultaneously making assignment decisions and the number of selectable tasks. Unlike settings with a fixed joint action space, both $m_t$ and $n_t$ change throughout an episode. Tasks become assigned or completed, and agents become temporarily committed to movement or execution. STAT therefore induces a state-dependent combinatorial action space whose complexity generally decreases as tasks are assigned and completed.

\subsection{Conflict Resolution}

Multiple agents may select the same task at the same assignment timestep. STAT resolves these assignment conflicts retrospectively. Let
\[
S_j(t)=\{i\in\mathcal{N}: a_i(t)=\textsc{select}(j)\}
\]
be the set of agents that select task $j$ at timestep $t$. If $|S_j(t)|>1$, the retained agent is
\[
i^*(j,t)=\arg\min_{i\in S_j(t)} d(x_i(t),y_j),
\]
with ties broken deterministically by agent index. Agent $i^*(j,t)$ receives assignment $g_{i^*}(t)=j$, while all other agents in $S_j(t)\setminus\{i^*(j,t)\}$ default to \textsc{idle}. The selected task is then marked as \textsc{assigned} and is no longer available for future selection.

This conflict-resolution rule makes coordination failures explicit and measurable. In STAT, a conflict is not only reflected indirectly through lower reward, it is an observable process-level event that reveals redundant allocation. Agents that lose conflict resolution do not make progress during that timestep, making redundant assignment costly through lost opportunity and delayed task completion. This allows the benchmark to distinguish policies that achieve similar return but differ in how efficiently they distribute agents across tasks.

\subsection{Movement, Execution, and Abstractions}

After an agent receives a task assignment, it moves toward the selected task. For an agent in \textsc{move} mode, the position update is
\[
x_i(t+1)
=
x_i(t)
+
\min\{v,d(x_i(t),y_{g_i(t)})\}
\frac{y_{g_i(t)}-x_i(t)}{d(x_i(t),y_{g_i(t)})},
\]
where $d(\cdot,\cdot)$ denotes Euclidean distance. When the agent reaches the assigned task location, it enters \textsc{executing}. After $K$ execution timesteps, the task is marked \textsc{completed}, and the agent becomes available to select another task if any remain. 

STAT abstracts away explicit agent-agent collision dynamics and low-level path-planning constraints. Agents move toward their assigned tasks without needing to solve a separate pathfinding problem. This abstraction prevents collision avoidance or complex navigation from becoming the dominant source of difficulty. The benchmark instead isolates spatial task-assignment coordination under combinatorial scaling.

\subsection{Reward Function}

We use a fixed reward function across all STAT configurations to provide a consistent task objective as problem complexity scales. Each agent $i\in\mathcal{N}$ receives an individual reward $\mathcal{R}_i(t)$ at time $t$, and the total team reward is
\[
\mathcal{R}_{\mathrm{total}}(t) = \sum_{i \in \mathcal{N}} \mathcal{R}_i(t).
\]

At each timestep, an agent receives a step penalty
\(
\mathcal{R}_i(t) = -\lambda_{\mathrm{step}},
\)
unless it completes a task. When an agent completes a task, it instead receives
\[
\mathcal{R}_i(t) =
\mathcal{R}^{b}(t)
\left(1 + \alpha T_{\mathrm{completed}}(t)\right),
\]
where $T_{\mathrm{completed}}(t)$ is the total number of tasks completed by time $t$, and $\alpha$ controls the progressive bonus for cumulative task completion. The base reward decays with elapsed time:
\[
\mathcal{R}^{b}(t)
=
R_0
-
\eta
\left\lfloor
\frac{\mathrm{steps}_t}{\beta}
\right\rfloor,
\]
where $R_0$ is the initial base reward, $\eta$ is the penalty applied at each decay interval, and $\beta$ is the interval length in timesteps. In our experiments, we set $R_0=30$, $\eta=0.5$, $\beta=10$, $\alpha=0.1$, and $\lambda_{\mathrm{step}}=1$, following prior work \cite{cardei2025factorizeddeepqnetworkcooperative}. We keep these reward parameters fixed across all benchmark settings so that methods optimize the same task objective.
% However, our main conclusions do not rely on reward alone; the benchmark analysis also reports process-level diagnostics that separately measure conflict, task-completion efficiency, and allocation quality.

\subsection{Environment Scope}

% An episode terminates when all tasks are completed or when the environment reaches the maximum episode length. 
STAT has the broad structure of a grid-based cooperative task-allocation problem, making it relevant to domains such as warehouse logistics, order fulfillment, delivery, victim tagging, and search and rescue \cite{Papadopoulos2025,krnjaic2024,cardei2025factorizeddeepqnetworkcooperative,zhu2023ofcourse}. The current benchmark intentionally uses homogeneous agents, homogeneous tasks, full observability, and simplified movement. This controlled scope is chosen to isolate structured combinatorial coordination without introducing additional confounds such as heterogeneous capabilities, partial observability, complex perception, collision avoidance, or domain-specific execution mechanics.

These simplifications are also natural extension points. STAT could be extended to heterogeneous agents by varying speed, specialization, or sensing capabilities; to heterogeneous tasks by varying execution time, priority, or completion requirements; and to partially observable settings by restricting each agent's access to the global state. 
In this work, we keep these factors fixed so that STAT isolates assignment
coordination and supports controlled analysis of how coordination processes
change as agent count, task count, and spatial extent are varied.

\begin{table*}[h]
\centering
\small
\renewcommand{\arraystretch}{1.18}
\caption{Overview of the cooperative environments and experimental variations used in our evaluation.}
\label{tab:environment_overview}

\begin{tabularx}{\textwidth}{@{}l c c X X@{}}
\toprule
\textbf{Environment}
& \textbf{Observability}
& \textbf{Agents}
& \textbf{Coordination Mechanism}
& \textbf{Studied Variation} \\
\midrule

STAT
& Full
& 3, 5, 9
& Commitment-constrained task assignment with directly observable assignment conflicts
& Agent count, task count, and spatial extent \\

LBF~\cite{christianos2020shared}
& Full / Partial
& 2, 3
& Capability matching and synchronized loading
& Agent count and observability \\

RWARE~\cite{papoudakis2021benchmarking}
& Partial
& 2, 4
& Multi-agent navigation, movement interference, congestion, and shelf delivery
& Agent count and warehouse layout \\

\bottomrule
\end{tabularx}
\end{table*}

% \begin{table*}[h]
% \centering
% \caption{Characteristics of representative cooperative MARL benchmarks relevant
% to process-aware evaluation.}
% \label{tab:benchmark_comparison}
% \resizebox{\textwidth}{!}{
% \begin{tabular}{lccccl}
% \toprule
% \textbf{Benchmark} &
% \textbf{Systematic Scale} &
% \textbf{Partial} &
% \textbf{Explicit Joint} &
% \textbf{Controlled} &
% \textbf{Primary Coordination Mechanism} \\
% &
% \textbf{Variation} &
% \textbf{Observability} &
% \textbf{Interaction} &
% \textbf{Isolation} &
% \\
% \midrule

% SMAC \cite{samvelyan2019smac}
% & \pmark & \cmark & \cmark & -- &
% Decentralized combat coordination \\

% MPE \cite{lowe2017}
% & \pmark & \pmark & \cmark & \pmark &
% Spatial coordination and communication \\

% LBF \cite{christianos2020shared}
% & \cmark & \pmark & \cmark & -- &
% Capability matching and synchronized loading \\

% Overcooked \cite{Carroll2019}
% & \pmark & \pmark & \cmark & -- &
% Collaborative planning and division of labor \\

% RWARE \cite{papoudakis2021benchmarking}
% & \cmark & \cmark & \cmark & -- &
% Navigation, congestion, pickup, and delivery \\

% \textbf{STAT (ours)}
% & \cmark & -- & \cmark & \cmark &
% Commitment-constrained task assignment \\

% \bottomrule
% \end{tabular}}
% \end{table*}

Table~\ref{tab:environment_overview} summarizes the three cooperative
environments used in our evaluation and the distinct coordination challenges
they expose. STAT provides controlled diagnostic depth by isolating
commitment-constrained task assignment and supporting systematic variation in
agent count, task count, and spatial extent. LBF introduces
capability-constrained synchronized cooperation under both full and partial
observability, while RWARE introduces partially observable multi-agent
navigation, movement interference, congestion, and sparse rewards. These environments provide complementary settings for evaluating whether the
process-aware framework remains informative across distinct coordination
mechanisms and experimental conditions.

% Table~\ref{tab:benchmark_comparison} compares STAT with commonly used cooperative MARL benchmarks along dimensions central to coordination-aware evaluation under scale. Existing benchmarks provide rich testbeds for cooperative behavior, while STAT is designed to complement them by isolating assignment coordination and exposing process-level coordination failures that may be hidden by return, success rate, win rate, or completion time alone.

\section{Established Environment Details}
\label{app:envs}
\subsection{Level-Based Foraging}

\begin{table}[h]
\centering
\small
\caption{LBF configurations used in the experiments.}
\label{tab:lbf_configs}
\begin{tabular}{lccc}
\toprule
\textbf{Setting} & \textbf{Agents} & \textbf{Observability} & \textbf{Episode Limit} \\
\midrule
2-agent full    & 2 & Full    & 50 \\
2-agent partial & 2 & Partial & 50 \\
3-agent full    & 3 & Full    & 50 \\
3-agent partial & 3 & Partial & 50 \\
\bottomrule
\end{tabular}
\end{table}

We use the Level-Based Foraging (LBF) environments through EPyMARL
\cite{papoudakis2021benchmarking} with the standard LBF environment
implementation. We evaluate two- and three-agent settings under both full and
partial observability. The exact environment identifiers and configuration
parameters used for each experiment are reported in
Table~\ref{tab:lbf_configs}. The corresponding environment IDs are
\texttt{Foraging-8x8-2p-2f-coop-v3},
\texttt{Foraging-2s-8x8-2p-2f-coop-v3},
\texttt{Foraging-10x10-3p-3f-v3}, and
\texttt{Foraging-2s-10x10-3p-3f-v3}, respectively.

Agents navigate a grid containing food items with associated levels. A food
item can be collected only when the combined level of participating agents is
sufficient for that item, creating capability-matching and synchronization
requirements. Agents use the environment's native movement, load, and NO-OP
actions. Episodes terminate according to the configured LBF episode limit or
when the task terminates naturally.

For all LBF experiments, we use a cooperative team reward by summing individual
environment rewards through the EPyMARL interface. We do not modify the native
transition, observation, or reward mechanics. Our only modifications are non-invasive logging hooks used to record movement,
loading, synchronization, and participation events for the process diagnostics
defined below; the native transition, observation, termination, and reward
mechanics are unchanged. Random seeds control both learning and
environment stochasticity.

\subsubsection{Process-Aware Metrics}
\label{app:lbf_metrics}

The LBF logging wrapper records process events at the food-target level rather
than treating each individual agent action as a separate coordination event.

\paragraph{Load Events.}
For each food item, a load event is counted once at a timestep when at least
one adjacent agent selects the \textsc{load} action. Thus, the denominator for
load-based rates is the number of food-target load events rather than the
number of individual agent \textsc{load} actions.

\paragraph{Failed-Load Rate.}
A failed load is a food-target load event for which the agents adjacent to the
food that select \textsc{load} do not jointly meet the food's level
requirement. We compute
\[
\mathrm{FailedLoadRate}
=
\frac{N_{\mathrm{failed\ load}}}
     {N_{\mathrm{load\ events}}}.
\]
This measures unsuccessful capability-constrained loading relative to the
number of opportunities in which agents attempted to collect a food item.

\paragraph{Successful Collection and Throughput.}
We report the number of food items actually removed from the environment,
\(N_{\mathrm{collected}}\), as the number of successfully collected foods.
Collection throughput is
\[
\rho_{\mathrm{LBF}}
=
\frac{N_{\mathrm{collected}}}{H},
\]
where \(H\) is the episode length. We additionally record \texttt{SuccessfulLoadEvents\_sum}, defined as the number of food-target load
events for which the summed level of the agents selecting \textsc{load}
satisfies the food's level requirement before the environment transition.

\paragraph{Synchronization Failures.}
A synchronization failure is a failed load event for which the full set of
agents adjacent to the targeted food collectively has sufficient level to
collect it, but the subset of adjacent agents that selects \textsc{load} does
not. We compute
\[
\mathrm{SynchronizationFailureRate}
=
\frac{N_{\mathrm{sync\ failure}}}
     {N_{\mathrm{load\ events}}}.
\]
This distinguishes failures caused by insufficient synchronized participation
from failures in which the locally available agent capacity is itself
insufficient.

\paragraph{Synchronization Delay.}
For each collected food item, we record the first timestep at which the agents
adjacent to that food collectively have sufficient level to satisfy its
collection requirement. Synchronization delay is the number of timesteps from
this first feasible timestep to the timestep at which the food is removed from
the environment. We average this quantity across collected food items for which
a feasible timestep was recorded.

\paragraph{Participation Balance.}
Let \(c_i\) denote the number of successfully collected-food events in which
agent \(i\) selects \textsc{load} as part of the successful loading group. We
define
\[
p_i = \frac{c_i}{\sum_j c_j},
\qquad
\mathrm{ParticipationBalance}
=
\frac{-\sum_i p_i \log p_i}{\log N},
\]
where \(N\) is the number of agents. The score ranges from 0 to 1, with higher
values indicating more evenly distributed successful participation. If there
is no successful participation in an episode, the score is set to zero.

\paragraph{Partner Diversity.}
For each agent \(i\), let \(\mathcal{P}_i\) denote the set of distinct
teammates with whom the agent participates in at least one successfully
collected-food event. We report the mean number of distinct partners per agent
and its normalized form,
\[
\mathrm{PartnerDiversityFrac}
=
\frac{\frac{1}{N}\sum_i |\mathcal{P}_i|}
     {N-1}.
\]
The normalized value ranges from 0 to 1, with higher values indicating that
successful cooperation is distributed across a broader set of teammate
pairings.

\subsection{Robotic Warehouse}

\begin{table}[t]
\centering
\small
\caption{RWARE configurations used in the experiments.}
\label{tab:rware_configs}
\begin{tabular}{lcc}
\toprule
\textbf{Setting} & \textbf{Agents} & \textbf{Episode Limit} \\
\midrule
\texttt{rware-tiny-2ag-v2}  & 2 & 500 \\
\texttt{rware-tiny-4ag-v2}  & 4 & 500 \\
\texttt{rware-small-4ag-v2} & 4 & 500 \\
\bottomrule
\end{tabular}
\end{table}

We use Robotic Warehouse (RWARE) through EPyMARL
\cite{papoudakis2021benchmarking} with the standard registered warehouse
environments. We evaluate
\texttt{rware-tiny-2ag-v2},
\texttt{rware-tiny-4ag-v2}, and
\texttt{rware-small-4ag-v2}.
The corresponding agent counts and episode limits are reported in
Table~\ref{tab:rware_configs}. All experiments use EPyMARL's Gymnasium
interface with \texttt{common\_reward=True},
\texttt{reward\_scalarisation=sum}, and
\texttt{env\_args.time\_limit=500}.

RWARE is partially observable and requires agents to navigate shared warehouse
aisles, retrieve requested shelves, and transport them to delivery locations.
Agents use the environment's native movement, rotation, shelf-interaction, and
NO-OP actions. Because the native environment provides individual rewards, we
use EPyMARL's sum scalarization to form a shared team reward. We otherwise
retain the native transition, observation, reward, and termination mechanics.

Our only modifications are non-invasive logging hooks used to record movement
interference, inactivity, shelf-delivery behavior, and agent contributions for
the process diagnostics defined below. Random seeds control both learning and
environment stochasticity.

\subsubsection{Process-Aware Metrics}

The RWARE diagnostics are derived from forward-movement requests, inactivity,
and delivery-reward events recorded during evaluation.

\paragraph{Canceled Movements.}
We record a movement request when an agent selects the native
\textsc{forward} action. A canceled movement is counted when the requested
forward target differs from the agent's current cell, but after the environment
resolves the transition the agent remains in its original cell. Thus, canceled
movements capture blocked forward progress under the environment's native
movement-resolution rules.

\paragraph{Canceled-Movement Rate.}
We normalize canceled movements by the number of forward requests,
\[
\mathrm{CanceledMovementRate}
=
\frac{N_{\mathrm{canceled\ forward}}}
     {N_{\mathrm{forward\ requests}}}.
\]
This metric captures how often attempted forward navigation fails to make
progress.

\paragraph{Shelf Deliveries.}
We infer delivery events from the per-agent reward vector before
common-reward aggregation. At each timestep, each agent receiving a positive
reward contributes one delivery event, and the resulting episode total is
logged as
\[
N_{\mathrm{delivered}}.
\]
This quantity corresponds to \texttt{ShelvesDelivered\_sum}. Because the count
is derived from positive per-agent rewards rather than directly from unique
shelf identifiers, it should be interpreted according to the reward convention
of the underlying RWARE environment.

\paragraph{Delivery Throughput.}
We normalize delivery events by episode length,
\[
\rho_{\mathrm{RWARE}}
=
\frac{N_{\mathrm{delivered}}}{H},
\]
where \(H\) is the number of timesteps in the episode.

\paragraph{NO-OP Rate.}
We measure inactivity as the fraction of agent-timestep actions that are the
native \textsc{noop} action,
\[
\mathrm{NoOpRate}
=
\frac{N_{\mathrm{NOOP}}}{HN},
\]
where \(N\) is the number of agents.

\paragraph{Steps Per Delivery.}
We log
\[
\mathrm{StepsPerDelivery}
=
\frac{H}{N_{\mathrm{delivered}}},
\]
using a safe-division convention that returns \(0\) when no delivery events
occur. We therefore interpret this metric jointly with delivery count and
throughput, rather than treating a zero value as indicating efficient task
completion.

\paragraph{Per-Agent Delivery Contribution.}
Let \(c_i\) denote the number of positive-reward delivery events attributed to
agent \(i\). The wrapper logs the per-agent counts
\(\{c_i\}_{i=1}^{N}\), their mean, and their coefficient of variation,
\[
\mathrm{CV}
=
\frac{\mathrm{std}(c_1,\ldots,c_N)}
     {\mathrm{mean}(c_1,\ldots,c_N)},
\]
with the coefficient of variation set to zero when the mean contribution is
zero. Lower CV indicates more even delivery contribution across agents to zero.

\section{Methods Evaluated}
\label{app:methods_evaluated}

\begin{table*}[t]
\centering
\small
\renewcommand{\arraystretch}{1.15}
\setlength{\tabcolsep}{5pt}
\caption{Overview of the MARL methods evaluated in this study. The methods
span centralized, centralized-training/decentralized-execution, and
independent learning approaches across value-based and actor--critic families.}
\label{tab:methods_overview}

\begin{tabularx}{\textwidth}{
@{}l
l
c
c
X
l@{}}
\toprule
\textbf{Method}
& \textbf{Method Family}
& \textbf{Train./Exec.}
& \textbf{Policy}
& \textbf{Coordination Structure}
& \textbf{Evaluated In} \\
\midrule

DQN \cite{mnih2015DQN}
& Value-based
& CTCE
& Off
& Centralized $Q$-learning over the joint action space
& STAT \\

FDQN \cite{cardei2025factorizeddeepqnetworkcooperative}
& Value-based
& CTCE
& Off
& Factorized centralized representation of the joint action space
& STAT \\

VDN \cite{Sunehag2017}
& Value decomposition
& CTDE
& Off
& Additive decomposition of the joint value into agent utilities
& STAT \\

QMIX \cite{rashid2018qmix}
& Value decomposition
& CTDE
& Off
& Monotonic mixing of individual agent utilities
& STAT, LBF \\

QTRAN \cite{son2019qtran}
& Value decomposition
& CTDE
& Off
& Joint value learning with constraints for decentralized execution
& STAT \\

IQL \cite{Tan1993}
& Independent value learning
& DTDE
& Off
& Independent per-agent $Q$-learning
& STAT, LBF \\

MAPPO \cite{yu2022surprising}
& Actor--critic
& CTDE
& On
& Decentralized actors with centralized value information during training
& STAT, LBF, RWARE \\

IPPO \cite{yu2022surprising}
& Actor--critic
& DTDE
& On
& Independent PPO optimization for each agent
& STAT, LBF, RWARE \\

MAA2C \cite{mnih2016asynchronous,papoudakis2021benchmarking}
& Actor--critic
& CTDE
& On
& Decentralized actors with a centralized state-value critic
& STAT \\

\bottomrule
\end{tabularx}
\end{table*}

Table~\ref{tab:methods_overview} summarizes the methods evaluated in this
study. The method set spans centralized joint-action learning, value
decomposition, independent learning, and actor-critic approaches with
different training and execution structures. This breadth allows us to examine
whether process-level coordination differences arise across substantially
different MARL paradigms rather than being specific to a single algorithm
family. Method coverage varies across environments according to benchmark
characteristics and the experimental design described in
Section~\ref{sec:experimental_setup}.

\paragraph{Centralized Training and Centralized Execution.}
In the CTCE paradigm, both learning and action selection are performed
centrally over the full multi-agent system. We evaluate DQN and FDQN.

\textbf{DQN.}
We extend Deep Q-Networks (DQN) \cite{mnih2015DQN} to a fully centralized
multi-agent setting by treating the global state and joint action as the input
and output of a single learner. The centralized Q-network directly estimates
the value of joint actions without factorization assumptions, but the output
space grows exponentially with the number of agents and available actions.

\textbf{FDQN.}
Factorized Deep Q-Networks (FDQN)
\cite{cardei2025factorizeddeepqnetworkcooperative} retain centralized training
and execution while factorizing the representation of the joint action space.
This reduces the computational burden of enumerating all joint actions while
preserving centralized reasoning over multi-agent decisions.

\paragraph{Centralized Training and Decentralized Execution.}
In the CTDE paradigm, agents may use centralized or joint information during
training while selecting actions independently at execution time. We evaluate
VDN, QMIX, QTRAN, MAPPO, and MAA2C.

\textbf{VDN.}
Value Decomposition Networks (VDN) \cite{Sunehag2017} represent the joint
action-value function as the sum of individual agent utilities. The joint
value is optimized using a temporal-difference objective, while each agent
selects actions from its local utility function at execution time.

\textbf{QMIX.}
QMIX \cite{rashid2018qmix} combines individual agent utilities through a
monotonic mixing network conditioned on global information. The monotonicity
constraint preserves consistency between the joint action-value maximization
and decentralized per-agent action selection.

\textbf{QTRAN.}
QTRAN \cite{son2019qtran} relaxes the monotonicity restriction of QMIX by
learning a centralized joint action-value function together with constraints
that preserve decentralized action selection. This allows a more expressive
representation of inter-agent dependencies.

\textbf{MAPPO.}
Multi-Agent Proximal Policy Optimization (MAPPO)
\cite{yu2022surprising} extends PPO to cooperative multi-agent settings using
decentralized actors and centralized value information during training. The
centralized critic can condition on global state information while each agent
executes its own policy independently.

\textbf{MAA2C.}
Multi-Agent A2C (MAA2C)
\cite{mnih2016asynchronous,papoudakis2021benchmarking} is an on-policy
actor-critic method with decentralized actors and a centralized state-value
critic. It provides a centralized-critic counterpart to MAPPO using the A2C
optimization framework.

\paragraph{Decentralized Training and Decentralized Execution.}
In the DTDE paradigm, agents learn and act independently using their own
observations and learning signals. We evaluate IQL and IPPO.

\textbf{IQL.}
Independent Q-Learning (IQL) \cite{Tan1993} trains a separate action-value
function for each agent using only that agent's local input. Each learner
updates independently and treats the behavior of the other agents as part of
the environment.

\textbf{IPPO.}
Independent Proximal Policy Optimization (IPPO)
\cite{yu2022surprising} applies PPO independently to each agent without a
centralized critic. Each agent learns its own policy from local information,
providing an independent actor-critic baseline for comparison with MAPPO.

\section{Implementation and Hyperparameter Tuning}
% \subsection{Hyperparameter Tuning}
\label{app:experimental_details}
% \paragraph{Implementation}
% We implement VDN, QMIX, QTRAN, and IQL using PyMARL \cite{samvelyan2019smac}, while FDQN follows implementation in \cite{cardei2025factorizeddeepqnetworkcooperative}, and DQN is implemented as its non-factorized centralized counterpart. 
\paragraph{Implementation.}
The value-based STAT baselines IQL, VDN, QMIX, and QTRAN are implemented
using a PyMARL-style codebase \cite{samvelyan2019smac}. FDQN follows the
factorized deep Q-network implementation of
\cite{cardei2025factorizeddeepqnetworkcooperative}, and DQN is implemented as
the corresponding non-factorized centralized joint-action baseline. MAPPO,
IPPO, and MAA2C are implemented using EPyMARL
\cite{papoudakis2021benchmarking}. LBF and RWARE are accessed through the
corresponding EPyMARL environment adapters, with non-invasive logging wrappers
added to record the process diagnostics defined in
Appendix~\ref{app:envs}.
Within each environment and configuration, methods use the same reward
specification, training budget, seed set, and evaluation protocol. Each
training run uses a single NVIDIA A100 GPU.
% The released artifact includes the
% exact environment and algorithm configuration files used for every experiment,
% together with the software versions and repository revisions required to
% reproduce the reported results.
\paragraph{STAT Value-Based Hyperparameter Selection.}

\begin{table}[h]
\centering
\small
\renewcommand{\arraystretch}{1.15}
\caption{Selected hyperparameters across representative benchmark settings. Small, Medium, and Large denote representative tuning settings used to select hyperparameters for the full benchmark suite. LR denotes learning rate, and $\epsilon$-decay denotes the fraction of training over which $\epsilon$ is decayed.}
\begin{tabular}{|l|cc|cc|cc|}
\hline
\multirow{2}{*}{\textbf{Algorithm}} 
& \multicolumn{2}{c|}{\textbf{Small}} 
& \multicolumn{2}{c|}{\textbf{Medium}} 
& \multicolumn{2}{c|}{\textbf{Large}} \\
\cline{2-7}
& \textbf{LR} & \textbf{$\epsilon$-decay} 
& \textbf{LR} & \textbf{$\epsilon$-decay} 
& \textbf{LR} & \textbf{$\epsilon$-decay} \\
\hline
DQN   & 0.0003 & 0.2 & -- & -- & -- & -- \\
\hline
FDQN  & 0.0003 & 0.2 & 0.001  & 0.2 & 0.0003 & 0.4 \\
\hline
VDN   & 0.001  & 0.2 & 0.001  & 0.2 & 0.001  & 0.2 \\
\hline
QMIX  & 0.001  & 0.2 & 0.001  & 0.4 & 0.001  & 0.4 \\
\hline
QTRAN & 0.001  & 0.2 & 0.001  & 0.4 & 0.001  & 0.4 \\
\hline
IQL   & 0.001  & 0.2 & 0.001  & 0.4 & 0.001  & 0.2 \\
\hline
\end{tabular}
\label{tab:hyperparams}
\end{table}

% We perform a lightweight hyperparameter search over learning rate and $\epsilon$-decay fraction on representative Small, Medium, and Large settings, then apply the selected configurations to the full benchmark suite. 
We perform a controlled, lightweight hyperparameter search over two optimization parameters: learning rate and $\epsilon$-decay fraction (Table \ref{tab:hyperparams}). For each algorithm, we evaluate learning rates of $3\times 10^{-4}$ and $10^{-3}$, and $\epsilon$-decay fractions of $0.2$ and $0.4$. Hyperparameter tuning is performed on three representative problem scales: Small, Medium, and Large. The hyperparameters selected on the Small setting are used for all 3-agent
STAT experiments. Medium-setting hyperparameters are used for 5-agent
experiments with environment sizes below $50\times30$, while Large-setting
hyperparameters are used for all $50\times30$ experiments. DQN is evaluated
only on the first three 3-agent configurations because its centralized
joint-action output becomes computationally infeasible at larger scales. This
scheme adapts hyperparameters to broad problem scale without tuning separately
for every configuration. For each algorithm and representative setting, we select the hyperparameter configuration that maximizes mean evaluation return over the final 20\% of evaluation checkpoints, averaged across training seeds. This criterion emphasizes stable late-stage evaluation performance rather than transient peaks during training. When multiple configurations produce similar final return, we break ties using average return over the full training trajectory, favoring stability and sample efficiency.

The hyperparameters selected from the Small setting are used for experiments with 3 agents, those selected from the Medium setting are used for experiments with 5 agents and environment sizes smaller than $50 \times 30$, and those selected from the Large setting are used for experiments with environment size $50 \times 30$. For DQN, it was only computationally feasible to run the first three experiments with 3 agents. This strategy adapts hyperparameters to broad changes in problem scale while avoiding per-configuration overfitting and maintaining a consistent evaluation protocol across the benchmark.

\paragraph{EPyMARL Hyperparameters.}
For MAPPO, IPPO, and MAA2C, we use the corresponding EPyMARL algorithm
configurations as the starting point, without per-environment hyperparameter
search. Unless otherwise stated, we keep the EPyMARL optimizer, network,
entropy, critic, and PPO/A2C update settings fixed across environments, and
vary only the experimental factors required by the study: environment
configuration, training budget, random seed, evaluation schedule, and, for
RWARE, the cooperative team-reward scalarization. Exact configurations used in each environment are reported in
Appendix~\ref{app:experimental_details}.

\section{Full Results in STAT}
\label{app:stat_results}

\subsection{Additional Outcome-Similarity Analysis}
\label{app:rq1_additional}
\begin{table}[h]
\centering
\small
\renewcommand{\arraystretch}{1.15}
\caption{Magnitude of process differences among outcome-similar STAT policy
pairs. Process gaps are computed as the ratio of the larger to smaller mean
value for assignment conflicts or forced idle. A pair is counted when at least
one of the two diagnostics reaches the specified threshold.}
\begin{tabular}{c|c|c|c|c}
\hline
\textbf{Return-Difference}
& \textbf{Outcome-Similar}
& $\mathbf{\geq 2\times}$
& $\mathbf{\geq 3\times}$
& $\mathbf{\geq 5\times}$ \\
\textbf{Threshold} & \textbf{Pairs} & & & \\
\hline
1\% & 121 & 38  & 26 & 9  \\
2\% & 199 & 80  & 47 & 20 \\
3\% & 244 & 103 & 65 & 27 \\
\hline
\end{tabular}
\label{tab:rq1_gap_magnitude}
\end{table}

The main text reports how often outcome-similar policy pairs differ by at
least $2\times$ in assignment conflicts or forced idle. To characterize the
magnitude of these hidden process differences more fully,
Table~\ref{tab:rq1_gap_magnitude} reports increasingly stringent process-gap
thresholds. Even under the strictest 1\% return-similarity criterion, 26 pairs
differ by at least $3\times$ in one or both diagnostics and 9 differ by at
least $5\times$.

\subsection{Additional STAT Results}
\label{app:full_env_results}
We report full final-policy results for all STAT configurations and methods.
Table~\ref{tab:stat_final_episodic_return} reports episodic return,
Table~\ref{tab:stat_final_conflicts_per_decision_opportunity} reports
conflicts per decision opportunity,
Table~\ref{tab:stat_final_task_throughput} reports task throughput,
Table~\ref{tab:stat_final_per_agent_assignment_diversity} reports per-agent
assignment diversity, and
Table~\ref{tab:stat_final_forced_idle_per_decision_opportunity} reports forced
idle per decision opportunity. These metrics correspond to the outcome and
process diagnostics shown in the learning curves in
Figures~\ref{fig:baseline_graphs} and~\ref{fig:extreme_graphs}.

Values are reported as mean $\pm$ 95\% confidence interval across five seeds.
For ease of comparison, bold indicates the best value according to the direction of each metric, and $^{\dagger}$ indicates methods that are not
significantly different from that value under Welch's two-sample $t$-test at
$\alpha=0.05$~\cite{welch1947generalization}. Higher episodic return and task
throughput are preferred, while lower conflicts and forced idle indicate less
coordination interference. Assignment diversity is interpreted jointly with
the other diagnostics rather than as a standalone measure of coordination
quality.

The full results reinforce that aggregate task performance and coordination
processes do not induce the same method ranking. In several configurations,
MAPPO and MAA2C achieve among the highest episodic returns and task throughput,
while methods with lower return can exhibit fewer assignment conflicts or less
forced idle. Conversely, low interference does not necessarily correspond to
high task performance or broad division of labor. These patterns show that no
single process diagnostic serves as a substitute for return and motivate
interpreting interference, task completion, participation, and coordination
efficiency jointly.
% The tables also provide the absolute metric values underlying the scaling analysis (Section \ref{sec:benchmark_analysis}), making it possible to inspect whether changes in return are associated with redundant assignment, reduced allocation breadth, or lower task-completion efficiency.

\begin{figure*}[h]
    \centering
\includegraphics[width=\linewidth]{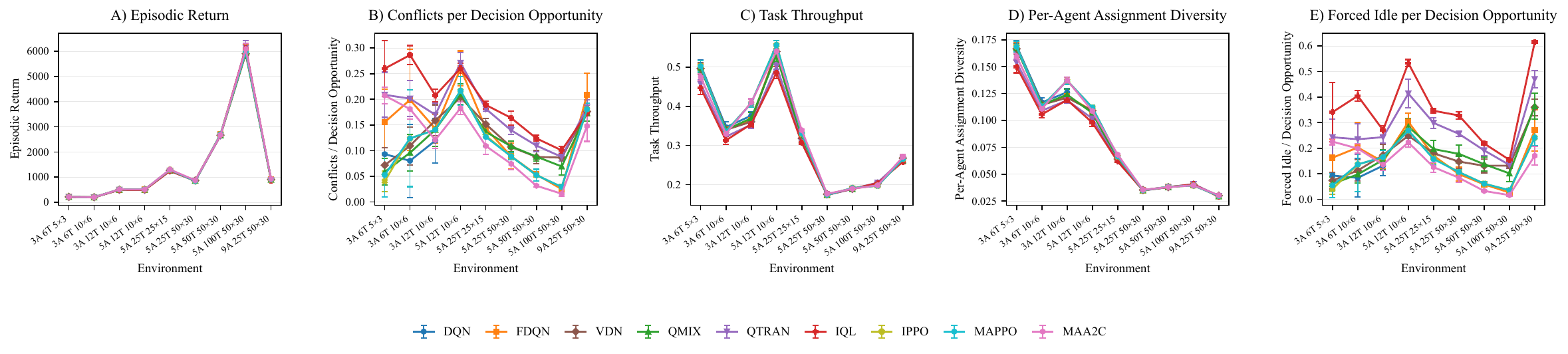}
\caption{Final performance across all STAT configurations. Points show means
across five seeds with 95\% confidence intervals. Alongside episodic return,
we report conflicts per decision opportunity, task throughput, per-agent
assignment diversity, and forced idle per decision opportunity. The resulting
profiles illustrate that strong aggregate performance does not necessarily
correspond to uniformly favorable values across all coordination processes.}
    \label{fig:summary_fig}
\end{figure*}

\begin{table*}[h]
\centering
\caption{Final episodic return over 5 seeds, reported as mean $\pm$ 95\% CI. Higher is better. Bold indicates the best method and $^{\dagger}$ indicates methods not significantly different from the best under Welch's two-sample $t$-test at $\alpha = 0.05$.}
\label{tab:stat_final_episodic_return}
\begin{adjustbox}{width=1\textwidth}
\begin{tabular}{lccccccccc}
\toprule
\textbf{Training Paradigm} & \multicolumn{2}{c}{\textbf{CTCE}} & \multicolumn{5}{c}{\textbf{CTDE}} & \multicolumn{2}{c}{\textbf{DTDE}} \\
\cmidrule(lr){2-3} \cmidrule(lr){4-8} \cmidrule(lr){9-10}
\textbf{Environments/Methods} & \textbf{DQN} & \textbf{FDQN} & \textbf{VDN} & \textbf{QMIX} & \textbf{QTRAN} & \textbf{MAPPO} & \textbf{MAA2C} & \textbf{IQL} & \textbf{IPPO} \\
\midrule
\textbf{3A-6T-5x3} & 215.05 $\pm$ 0.7$^{\dagger}$ & 214.97 $\pm$ 0.7$^{\dagger}$ & 214.59 $\pm$ 1.1$^{\dagger}$ & 215.06 $\pm$ 0.4$^{\dagger}$ & 212.31 $\pm$ 1.0 & \textbf{215.23 $\pm$ 0.9} & 213.81 $\pm$ 1.2 & 211.73 $\pm$ 1.4 & 214.86 $\pm$ 0.9$^{\dagger}$ \\
\textbf{3A-6T-10x6} & \textbf{199.94 $\pm$ 1.4} & 199.87 $\pm$ 0.7$^{\dagger}$ & 199.76 $\pm$ 0.8$^{\dagger}$ & 199.61 $\pm$ 0.8$^{\dagger}$ & 196.90 $\pm$ 1.5 & 199.64 $\pm$ 0.7$^{\dagger}$ & 199.14 $\pm$ 1.9$^{\dagger}$ & 195.52 $\pm$ 0.7 & 199.64 $\pm$ 0.7$^{\dagger}$ \\
\textbf{3A-12T-10x6} & 504.20 $\pm$ 2.5 & 499.72 $\pm$ 2.2 & 499.14 $\pm$ 2.6 & 502.07 $\pm$ 3.7 & 495.87 $\pm$ 3.4 & 512.93 $\pm$ 1.2$^{\dagger}$ & \textbf{513.22 $\pm$ 1.5} & 496.29 $\pm$ 1.9 & 512.93 $\pm$ 1.2$^{\dagger}$ \\
\textbf{5A-12T-10x6} & -- & 503.68 $\pm$ 2.5 & 501.31 $\pm$ 3.2 & 500.11 $\pm$ 2.2 & 492.66 $\pm$ 1.8 & \textbf{507.39 $\pm$ 2.7} & 506.78 $\pm$ 1.4$^{\dagger}$ & 487.81 $\pm$ 3.0 & 507.39 $\pm$ 2.7$^{\dagger}$ \\
\textbf{5A-25T-25x15} & -- & 1263.87 $\pm$ 14.2 & 1266.00 $\pm$ 5.3 & 1261.98 $\pm$ 8.5 & 1248.16 $\pm$ 7.9 & 1290.53 $\pm$ 5.1 & \textbf{1303.45 $\pm$ 9.7} & 1245.52 $\pm$ 5.4 & 1288.61 $\pm$ 9.9 \\
\textbf{5A-25T-50x30} & -- & 857.58 $\pm$ 32.5$^{\dagger}$ & 866.89 $\pm$ 26.3$^{\dagger}$ & 856.06 $\pm$ 17.1$^{\dagger}$ & 853.16 $\pm$ 16.9$^{\dagger}$ & 845.74 $\pm$ 18.5$^{\dagger}$ & \textbf{872.14 $\pm$ 30.7} & 854.01 $\pm$ 9.0$^{\dagger}$ & 845.74 $\pm$ 18.5$^{\dagger}$ \\
\textbf{5A-50T-50x30} & -- & 2697.84 $\pm$ 19.6$^{\dagger}$ & 2666.88 $\pm$ 46.6$^{\dagger}$ & 2664.23 $\pm$ 66.7$^{\dagger}$ & 2648.79 $\pm$ 52.4$^{\dagger}$ & \textbf{2700.71 $\pm$ 47.9} & 2671.79 $\pm$ 49.6$^{\dagger}$ & 2674.91 $\pm$ 35.8$^{\dagger}$ & 2700.71 $\pm$ 47.9$^{\dagger}$ \\
\textbf{5A-100T-50x30} & -- & 6075.01 $\pm$ 198.8$^{\dagger}$ & 5900.39 $\pm$ 194.1$^{\dagger}$ & 5934.09 $\pm$ 207.3$^{\dagger}$ & \textbf{6144.00 $\pm$ 268.8} & 5974.01 $\pm$ 128.0$^{\dagger}$ & 6090.18 $\pm$ 223.2$^{\dagger}$ & 6106.37 $\pm$ 206.3$^{\dagger}$ & 5974.01 $\pm$ 128.0$^{\dagger}$ \\
\textbf{9A-25T-50x30} & -- & 909.51 $\pm$ 28.6$^{\dagger}$ & 909.08 $\pm$ 21.1$^{\dagger}$ & 899.69 $\pm$ 22.8 & 888.71 $\pm$ 25.6 & 907.66 $\pm$ 30.8$^{\dagger}$ & \textbf{928.70 $\pm$ 14.9} & 872.16 $\pm$ 11.6 & 907.66 $\pm$ 30.8$^{\dagger}$ \\
\bottomrule
\end{tabular}
\end{adjustbox}
\end{table*}

\begin{table*}[h]
\centering
\caption{Final conflicts / decision opportunity over 5 seeds, reported as mean $\pm$ 95\% CI. Lower is better. Bold indicates the best method and $^{\dagger}$ indicates methods not significantly different from the best under Welch's two-sample $t$-test at $\alpha = 0.05$.}
\label{tab:stat_final_conflicts_per_decision_opportunity}
\begin{adjustbox}{width=1\textwidth}
\begin{tabular}{lccccccccc}
\toprule
\textbf{Training Paradigm} & \multicolumn{2}{c}{\textbf{CTCE}} & \multicolumn{5}{c}{\textbf{CTDE}} & \multicolumn{2}{c}{\textbf{DTDE}} \\
\cmidrule(lr){2-3} \cmidrule(lr){4-8} \cmidrule(lr){9-10}
\textbf{Environments/Methods} & \textbf{DQN} & \textbf{FDQN} & \textbf{VDN} & \textbf{QMIX} & \textbf{QTRAN} & \textbf{MAPPO} & \textbf{MAA2C} & \textbf{IQL} & \textbf{IPPO} \\
\midrule
\textbf{3A-6T-5x3} & 0.093 $\pm$ 0.073$^{\dagger}$ & 0.156 $\pm$ 0.064 & 0.072 $\pm$ 0.034$^{\dagger}$ & 0.059 $\pm$ 0.026$^{\dagger}$ & 0.209 $\pm$ 0.044 & 0.053 $\pm$ 0.042$^{\dagger}$ & 0.207 $\pm$ 0.016 & 0.260 $\pm$ 0.054 & \textbf{0.040 $\pm$ 0.020} \\
\textbf{3A-6T-10x6} & \textbf{0.080 $\pm$ 0.071} & 0.199 $\pm$ 0.099 & 0.110 $\pm$ 0.037$^{\dagger}$ & 0.096 $\pm$ 0.036$^{\dagger}$ & 0.201 $\pm$ 0.035 & 0.124 $\pm$ 0.094$^{\dagger}$ & 0.181 $\pm$ 0.020 & 0.286 $\pm$ 0.018 & 0.124 $\pm$ 0.094$^{\dagger}$ \\
\textbf{3A-12T-10x6} & \textbf{0.120 $\pm$ 0.043} & 0.143 $\pm$ 0.027$^{\dagger}$ & 0.159 $\pm$ 0.030$^{\dagger}$ & 0.141 $\pm$ 0.032$^{\dagger}$ & 0.170 $\pm$ 0.023 & 0.141 $\pm$ 0.020$^{\dagger}$ & 0.121 $\pm$ 0.017$^{\dagger}$ & 0.208 $\pm$ 0.012 & 0.141 $\pm$ 0.020$^{\dagger}$ \\
\textbf{5A-12T-10x6} & -- & 0.260 $\pm$ 0.034 & 0.204 $\pm$ 0.026$^{\dagger}$ & 0.210 $\pm$ 0.020 & 0.271 $\pm$ 0.020 & 0.217 $\pm$ 0.028 & \textbf{0.183 $\pm$ 0.012} & 0.260 $\pm$ 0.010 & 0.217 $\pm$ 0.028 \\
\textbf{5A-25T-25x15} & -- & 0.143 $\pm$ 0.008 & 0.152 $\pm$ 0.012 & 0.137 $\pm$ 0.011 & 0.182 $\pm$ 0.005 & 0.127 $\pm$ 0.019$^{\dagger}$ & \textbf{0.109 $\pm$ 0.016} & 0.189 $\pm$ 0.007 & 0.129 $\pm$ 0.019$^{\dagger}$ \\
\textbf{5A-25T-50x30} & -- & 0.089 $\pm$ 0.025$^{\dagger}$ & 0.107 $\pm$ 0.011 & 0.108 $\pm$ 0.011 & 0.140 $\pm$ 0.008 & 0.089 $\pm$ 0.007 & \textbf{0.074 $\pm$ 0.012} & 0.164 $\pm$ 0.014 & 0.089 $\pm$ 0.007 \\
\textbf{5A-50T-50x30} & -- & 0.054 $\pm$ 0.005 & 0.088 $\pm$ 0.013 & 0.088 $\pm$ 0.009 & 0.110 $\pm$ 0.012 & 0.052 $\pm$ 0.011 & \textbf{0.032 $\pm$ 0.002} & 0.124 $\pm$ 0.006 & 0.052 $\pm$ 0.011 \\
\textbf{5A-100T-50x30} & -- & 0.025 $\pm$ 0.006 & 0.086 $\pm$ 0.007 & 0.069 $\pm$ 0.017 & 0.088 $\pm$ 0.008 & 0.029 $\pm$ 0.005 & \textbf{0.016 $\pm$ 0.005} & 0.101 $\pm$ 0.007 & 0.029 $\pm$ 0.005 \\
\textbf{9A-25T-50x30} & -- & 0.209 $\pm$ 0.041 & 0.176 $\pm$ 0.008$^{\dagger}$ & 0.173 $\pm$ 0.015$^{\dagger}$ & 0.185 $\pm$ 0.014 & 0.181 $\pm$ 0.012 & \textbf{0.148 $\pm$ 0.030} & 0.173 $\pm$ 0.005$^{\dagger}$ & 0.181 $\pm$ 0.012 \\
\bottomrule
\end{tabular}
\end{adjustbox}
\end{table*}

\begin{table*}[h]
\centering
\caption{Final task throughput over 5 seeds, reported as mean $\pm$ 95\% CI. Higher is better. Bold indicates the best method and $^{\dagger}$ indicates methods not significantly different from the best under Welch's two-sample $t$-test at $\alpha = 0.05$.}
\label{tab:stat_final_task_throughput}
\begin{adjustbox}{width=1\textwidth}
\begin{tabular}{lccccccccc}
\toprule
\textbf{Training Paradigm} & \multicolumn{2}{c}{\textbf{CTCE}} & \multicolumn{5}{c}{\textbf{CTDE}} & \multicolumn{2}{c}{\textbf{DTDE}} \\
\cmidrule(lr){2-3} \cmidrule(lr){4-8} \cmidrule(lr){9-10}
\textbf{Environments/Methods} & \textbf{DQN} & \textbf{FDQN} & \textbf{VDN} & \textbf{QMIX} & \textbf{QTRAN} & \textbf{MAPPO} & \textbf{MAA2C} & \textbf{IQL} & \textbf{IPPO} \\
\midrule
\textbf{3A-6T-5x3} & \textbf{0.505 $\pm$ 0.013} & 0.501 $\pm$ 0.015$^{\dagger}$ & 0.497 $\pm$ 0.015$^{\dagger}$ & 0.498 $\pm$ 0.011$^{\dagger}$ & 0.457 $\pm$ 0.011 & 0.502 $\pm$ 0.015$^{\dagger}$ & 0.475 $\pm$ 0.008 & 0.447 $\pm$ 0.017 & 0.503 $\pm$ 0.015$^{\dagger}$ \\
\textbf{3A-6T-10x6} & \textbf{0.347 $\pm$ 0.013} & 0.340 $\pm$ 0.007$^{\dagger}$ & 0.340 $\pm$ 0.008$^{\dagger}$ & 0.340 $\pm$ 0.007$^{\dagger}$ & 0.322 $\pm$ 0.011 & 0.335 $\pm$ 0.007$^{\dagger}$ & 0.329 $\pm$ 0.015 & 0.313 $\pm$ 0.010 & 0.335 $\pm$ 0.007$^{\dagger}$ \\
\textbf{3A-12T-10x6} & 0.376 $\pm$ 0.010 & 0.364 $\pm$ 0.008 & 0.361 $\pm$ 0.006 & 0.369 $\pm$ 0.015 & 0.353 $\pm$ 0.010 & 0.408 $\pm$ 0.010$^{\dagger}$ & \textbf{0.410 $\pm$ 0.009} & 0.354 $\pm$ 0.007 & 0.408 $\pm$ 0.010$^{\dagger}$ \\
\textbf{5A-12T-10x6} & -- & 0.541 $\pm$ 0.019$^{\dagger}$ & 0.540 $\pm$ 0.014 & 0.530 $\pm$ 0.015 & 0.503 $\pm$ 0.009 & \textbf{0.557 $\pm$ 0.010} & 0.541 $\pm$ 0.015 & 0.486 $\pm$ 0.015 & 0.557 $\pm$ 0.010$^{\dagger}$ \\
\textbf{5A-25T-25x15} & -- & 0.318 $\pm$ 0.008 & 0.318 $\pm$ 0.003 & 0.314 $\pm$ 0.005 & 0.309 $\pm$ 0.005 & 0.330 $\pm$ 0.003 & \textbf{0.339 $\pm$ 0.004} & 0.307 $\pm$ 0.004 & 0.329 $\pm$ 0.005 \\
\textbf{5A-25T-50x30} & -- & 0.174 $\pm$ 0.007$^{\dagger}$ & 0.175 $\pm$ 0.006$^{\dagger}$ & 0.174 $\pm$ 0.003$^{\dagger}$ & 0.173 $\pm$ 0.002$^{\dagger}$ & 0.174 $\pm$ 0.004$^{\dagger}$ & \textbf{0.176 $\pm$ 0.005} & 0.174 $\pm$ 0.004$^{\dagger}$ & 0.174 $\pm$ 0.004$^{\dagger}$ \\
\textbf{5A-50T-50x30} & -- & 0.190 $\pm$ 0.002$^{\dagger}$ & 0.189 $\pm$ 0.003$^{\dagger}$ & 0.189 $\pm$ 0.004$^{\dagger}$ & 0.189 $\pm$ 0.003$^{\dagger}$ & \textbf{0.191 $\pm$ 0.003} & 0.189 $\pm$ 0.002$^{\dagger}$ & 0.189 $\pm$ 0.003$^{\dagger}$ & 0.191 $\pm$ 0.003$^{\dagger}$ \\
\textbf{5A-100T-50x30} & -- & 0.198 $\pm$ 0.003 & 0.202 $\pm$ 0.001$^{\dagger}$ & 0.203 $\pm$ 0.002$^{\dagger}$ & \textbf{0.205 $\pm$ 0.003} & 0.197 $\pm$ 0.002 & 0.198 $\pm$ 0.003 & 0.203 $\pm$ 0.002$^{\dagger}$ & 0.197 $\pm$ 0.002 \\
\textbf{9A-25T-50x30} & -- & 0.266 $\pm$ 0.009$^{\dagger}$ & 0.268 $\pm$ 0.006$^{\dagger}$ & 0.262 $\pm$ 0.005 & 0.265 $\pm$ 0.009$^{\dagger}$ & 0.265 $\pm$ 0.006$^{\dagger}$ & \textbf{0.271 $\pm$ 0.006} & 0.258 $\pm$ 0.005 & 0.265 $\pm$ 0.006$^{\dagger}$ \\
\bottomrule
\end{tabular}
\end{adjustbox}
\end{table*}

\begin{table*}[h]
\centering
\caption{Final per-agent assignment diversity over 5 seeds, reported as mean $\pm$ 95\% CI. Higher is better. Bold indicates the best method and $^{\dagger}$ indicates methods not significantly different from the best under Welch's two-sample $t$-test at $\alpha = 0.05$.}
\label{tab:stat_final_per_agent_assignment_diversity}
\begin{adjustbox}{width=1\textwidth}
\begin{tabular}{lccccccccc}
\toprule
\textbf{Training Paradigm} & \multicolumn{2}{c}{\textbf{CTCE}} & \multicolumn{5}{c}{\textbf{CTDE}} & \multicolumn{2}{c}{\textbf{DTDE}} \\
\cmidrule(lr){2-3} \cmidrule(lr){4-8} \cmidrule(lr){9-10}
\textbf{Environments/Methods} & \textbf{DQN} & \textbf{FDQN} & \textbf{VDN} & \textbf{QMIX} & \textbf{QTRAN} & \textbf{MAPPO} & \textbf{MAA2C} & \textbf{IQL} & \textbf{IPPO} \\
\midrule
\textbf{3A-6T-5x3} & \textbf{0.170 $\pm$ 0.004} & 0.168 $\pm$ 0.005$^{\dagger}$ & 0.167 $\pm$ 0.005$^{\dagger}$ & 0.167 $\pm$ 0.004$^{\dagger}$ & 0.154 $\pm$ 0.003 & 0.169 $\pm$ 0.005$^{\dagger}$ & 0.160 $\pm$ 0.003 & 0.150 $\pm$ 0.006 & 0.169 $\pm$ 0.005$^{\dagger}$ \\
\textbf{3A-6T-10x6} & \textbf{0.117 $\pm$ 0.004} & 0.115 $\pm$ 0.002$^{\dagger}$ & 0.115 $\pm$ 0.003$^{\dagger}$ & 0.115 $\pm$ 0.002$^{\dagger}$ & 0.109 $\pm$ 0.003 & 0.113 $\pm$ 0.002$^{\dagger}$ & 0.111 $\pm$ 0.005 & 0.106 $\pm$ 0.003 & 0.113 $\pm$ 0.002$^{\dagger}$ \\
\textbf{3A-12T-10x6} & 0.126 $\pm$ 0.004 & 0.122 $\pm$ 0.002 & 0.121 $\pm$ 0.002 & 0.124 $\pm$ 0.005 & 0.119 $\pm$ 0.003 & 0.137 $\pm$ 0.003$^{\dagger}$ & \textbf{0.137 $\pm$ 0.003} & 0.119 $\pm$ 0.003 & 0.137 $\pm$ 0.003$^{\dagger}$ \\
\textbf{5A-12T-10x6} & -- & 0.109 $\pm$ 0.004$^{\dagger}$ & 0.109 $\pm$ 0.003$^{\dagger}$ & 0.107 $\pm$ 0.003 & 0.102 $\pm$ 0.002 & \textbf{0.112 $\pm$ 0.002} & 0.109 $\pm$ 0.003 & 0.098 $\pm$ 0.003 & 0.112 $\pm$ 0.002$^{\dagger}$ \\
\textbf{5A-25T-25x15} & -- & 0.064 $\pm$ 0.001 & 0.064 $\pm$ 0.001 & 0.063 $\pm$ 0.001 & 0.062 $\pm$ 0.001 & 0.067 $\pm$ 0.000 & \textbf{0.068 $\pm$ 0.001} & 0.062 $\pm$ 0.001 & 0.066 $\pm$ 0.001 \\
\textbf{5A-25T-50x30} & -- & 0.035 $\pm$ 0.001$^{\dagger}$ & 0.035 $\pm$ 0.001$^{\dagger}$ & 0.035 $\pm$ 0.001$^{\dagger}$ & 0.035 $\pm$ 0.000$^{\dagger}$ & 0.035 $\pm$ 0.001$^{\dagger}$ & \textbf{0.036 $\pm$ 0.001} & 0.035 $\pm$ 0.001$^{\dagger}$ & 0.035 $\pm$ 0.001$^{\dagger}$ \\
\textbf{5A-50T-50x30} & -- & 0.038 $\pm$ 0.000$^{\dagger}$ & 0.038 $\pm$ 0.001$^{\dagger}$ & 0.038 $\pm$ 0.001$^{\dagger}$ & 0.038 $\pm$ 0.001$^{\dagger}$ & \textbf{0.038 $\pm$ 0.001} & 0.038 $\pm$ 0.001$^{\dagger}$ & 0.038 $\pm$ 0.001$^{\dagger}$ & 0.038 $\pm$ 0.001$^{\dagger}$ \\
\textbf{5A-100T-50x30} & -- & 0.040 $\pm$ 0.001 & 0.040 $\pm$ 0.000$^{\dagger}$ & 0.040 $\pm$ 0.000$^{\dagger}$ & \textbf{0.041 $\pm$ 0.001} & 0.039 $\pm$ 0.000 & 0.040 $\pm$ 0.001 & 0.041 $\pm$ 0.000$^{\dagger}$ & 0.039 $\pm$ 0.000 \\
\textbf{9A-25T-50x30} & -- & 0.030 $\pm$ 0.001$^{\dagger}$ & 0.030 $\pm$ 0.001$^{\dagger}$ & 0.029 $\pm$ 0.001 & 0.030 $\pm$ 0.001$^{\dagger}$ & 0.030 $\pm$ 0.001$^{\dagger}$ & \textbf{0.030 $\pm$ 0.001} & 0.029 $\pm$ 0.001 & 0.030 $\pm$ 0.001$^{\dagger}$ \\
\bottomrule
\end{tabular}
\end{adjustbox}
\end{table*}

\begin{table*}[h]
\centering
\caption{Final forced idle / decision opportunity over 5 seeds, reported as mean $\pm$ 95\% CI. Lower is better. Bold indicates the best method and $^{\dagger}$ indicates methods not significantly different from the best under Welch's two-sample $t$-test at $\alpha = 0.05$.}
\label{tab:stat_final_forced_idle_per_decision_opportunity}
\begin{adjustbox}{width=1\textwidth}
\begin{tabular}{lccccccccc}
\toprule
\textbf{Training Paradigm} & \multicolumn{2}{c}{\textbf{CTCE}} & \multicolumn{5}{c}{\textbf{CTDE}} & \multicolumn{2}{c}{\textbf{DTDE}} \\
\cmidrule(lr){2-3} \cmidrule(lr){4-8} \cmidrule(lr){9-10}
\textbf{Environments/Methods} & \textbf{DQN} & \textbf{FDQN} & \textbf{VDN} & \textbf{QMIX} & \textbf{QTRAN} & \textbf{MAPPO} & \textbf{MAA2C} & \textbf{IQL} & \textbf{IPPO} \\
\midrule
\textbf{3A-6T-5x3} & 0.093 $\pm$ 0.073$^{\dagger}$ & 0.163 $\pm$ 0.073 & 0.074 $\pm$ 0.033 & 0.060 $\pm$ 0.027$^{\dagger}$ & 0.243 $\pm$ 0.072 & 0.054 $\pm$ 0.047$^{\dagger}$ & 0.226 $\pm$ 0.012 & 0.341 $\pm$ 0.117 & \textbf{0.040 $\pm$ 0.020} \\
\textbf{3A-6T-10x6} & \textbf{0.084 $\pm$ 0.076} & 0.203 $\pm$ 0.098 & 0.113 $\pm$ 0.041$^{\dagger}$ & 0.098 $\pm$ 0.033$^{\dagger}$ & 0.235 $\pm$ 0.060 & 0.137 $\pm$ 0.107$^{\dagger}$ & 0.199 $\pm$ 0.018 & 0.405 $\pm$ 0.021 & 0.137 $\pm$ 0.107$^{\dagger}$ \\
\textbf{3A-12T-10x6} & \textbf{0.132 $\pm$ 0.039} & 0.147 $\pm$ 0.027$^{\dagger}$ & 0.173 $\pm$ 0.043$^{\dagger}$ & 0.154 $\pm$ 0.039$^{\dagger}$ & 0.243 $\pm$ 0.023 & 0.165 $\pm$ 0.027$^{\dagger}$ & 0.136 $\pm$ 0.021$^{\dagger}$ & 0.271 $\pm$ 0.018 & 0.165 $\pm$ 0.027$^{\dagger}$ \\
\textbf{5A-12T-10x6} & -- & 0.303 $\pm$ 0.033 & 0.249 $\pm$ 0.034$^{\dagger}$ & 0.284 $\pm$ 0.033 & 0.414 $\pm$ 0.055 & 0.269 $\pm$ 0.031 & \textbf{0.223 $\pm$ 0.019} & 0.533 $\pm$ 0.014 & 0.269 $\pm$ 0.031 \\
\textbf{5A-25T-25x15} & -- & 0.171 $\pm$ 0.015 & 0.180 $\pm$ 0.014 & 0.197 $\pm$ 0.034 & 0.299 $\pm$ 0.021 & 0.158 $\pm$ 0.022 & \textbf{0.126 $\pm$ 0.020} & 0.346 $\pm$ 0.010 & 0.161 $\pm$ 0.021 \\
\textbf{5A-25T-50x30} & -- & 0.099 $\pm$ 0.026$^{\dagger}$ & 0.148 $\pm$ 0.026 & 0.178 $\pm$ 0.035 & 0.256 $\pm$ 0.010 & 0.107 $\pm$ 0.008 & \textbf{0.084 $\pm$ 0.018} & 0.328 $\pm$ 0.013 & 0.107 $\pm$ 0.008 \\
\textbf{5A-50T-50x30} & -- & 0.059 $\pm$ 0.008 & 0.131 $\pm$ 0.027 & 0.137 $\pm$ 0.027 & 0.192 $\pm$ 0.022 & 0.061 $\pm$ 0.007 & \textbf{0.033 $\pm$ 0.003} & 0.218 $\pm$ 0.006 & 0.061 $\pm$ 0.007 \\
\textbf{5A-100T-50x30} & -- & 0.027 $\pm$ 0.007 & 0.132 $\pm$ 0.010 & 0.102 $\pm$ 0.033 & 0.135 $\pm$ 0.011 & 0.036 $\pm$ 0.009 & \textbf{0.017 $\pm$ 0.005} & 0.154 $\pm$ 0.007 & 0.036 $\pm$ 0.009 \\
\textbf{9A-25T-50x30} & -- & 0.270 $\pm$ 0.080 & 0.360 $\pm$ 0.033 & 0.365 $\pm$ 0.051 & 0.470 $\pm$ 0.033 & 0.241 $\pm$ 0.031 & \textbf{0.171 $\pm$ 0.037} & 0.615 $\pm$ 0.005 & 0.241 $\pm$ 0.031 \\
\bottomrule
\end{tabular}
\end{adjustbox}
\end{table*}

\subsection{STAT Learning Curves} \label{app:learning_curves}
\begin{figure*}[h]
    \centering
\includegraphics[width=\textwidth]{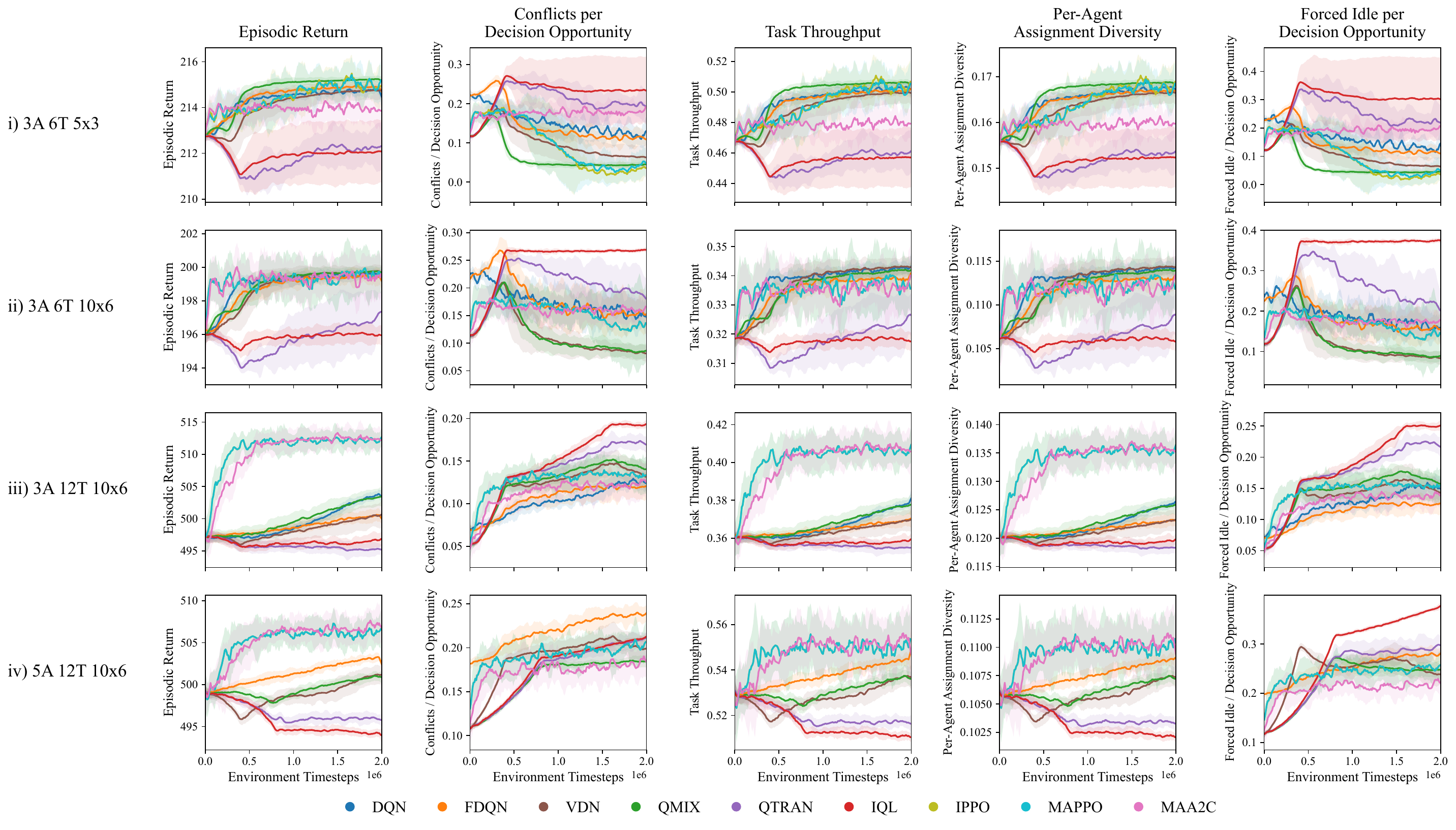}
\caption{Learning curves for the Baseline STAT configurations. Curves show
mean evaluation performance across five seeds, with shaded regions denoting
95\% confidence intervals. We report episodic return as the aggregate outcome
measure together with four process diagnostics: conflicts per decision
opportunity, task throughput, per-agent assignment diversity, and forced idle
per decision opportunity. These diagnostics capture complementary aspects of
assignment interference, task-completion efficiency, division of labor, and
the immediate cost of coordination failures.}
    \label{fig:baseline_graphs}
\end{figure*}

\begin{figure*}[h]
    \centering
\includegraphics[width=\textwidth]{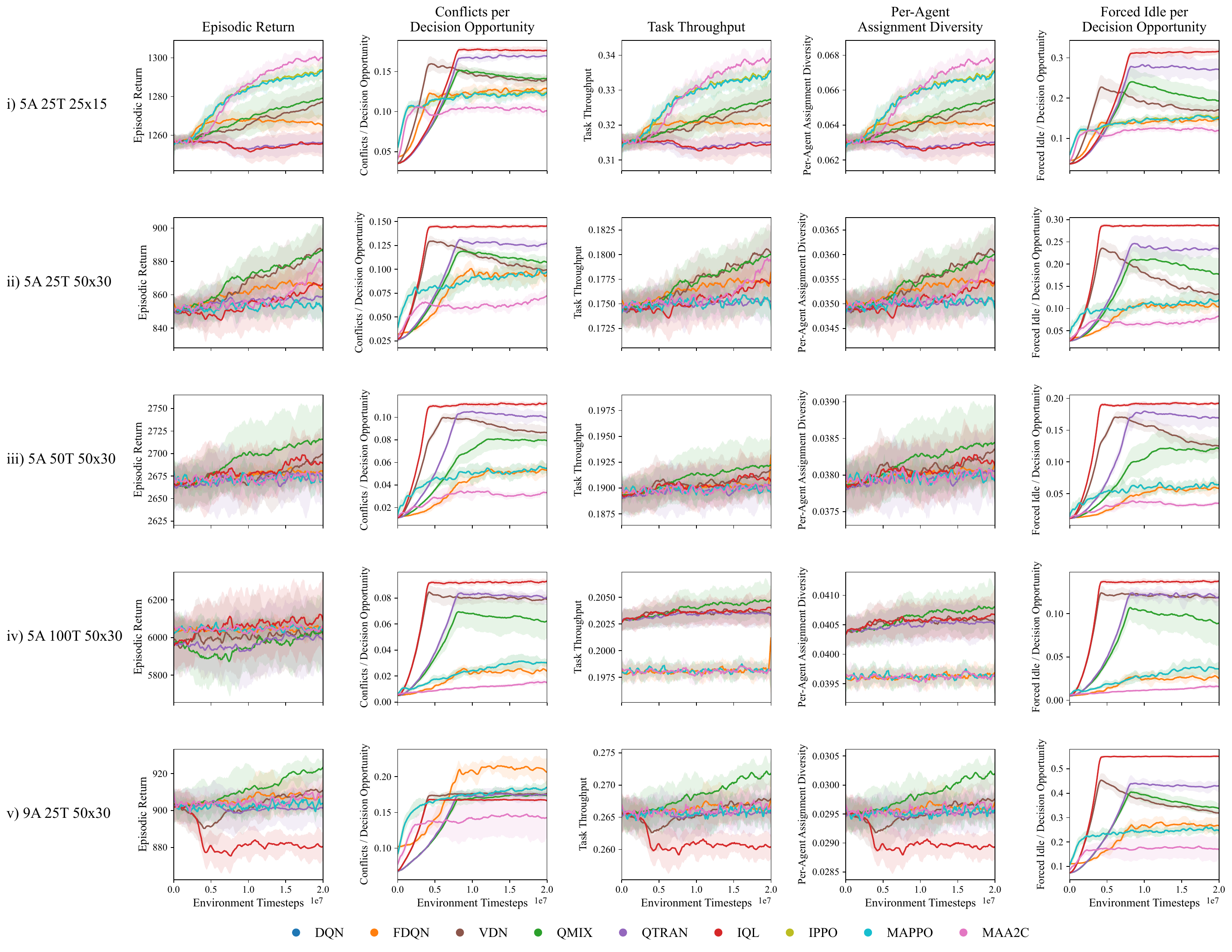}
\caption{Learning curves for the Extreme STAT configurations. Curves show
mean evaluation performance across five seeds, with shaded regions denoting
95\% confidence intervals.}
    \label{fig:extreme_graphs}
\end{figure*}

For each STAT configuration, we plot evaluation return together with the four
process-level diagnostics over training. Baseline configurations are shown in
Figure~\ref{fig:baseline_graphs}, and Extreme configurations are shown in
Figure~\ref{fig:extreme_graphs}. The curves illustrate how task performance,
assignment interference, task throughput, division of labor, and
conflict-induced inactivity evolve as policies are learned.

Across the Baseline configurations, improvements in return are often
accompanied by higher throughput and assignment diversity, but not necessarily
by lower interference. In particular, some high-return policies retain
non-negligible conflicts and forced idle, while other methods achieve lower
interference without attaining the highest return. This further illustrates
that individual process metrics do not provide interchangeable measures of
policy quality and should be interpreted jointly. For example, MAPPO and MAA2C often achieve strong return and throughput while
retaining non-negligible conflict and forced-idle levels.
% , illustrating that
% different process metrics may not improve similar to aggregate performance.

The Extreme configurations amplify these differences. As task sets, spatial
extent, or team size increase, methods exhibit increasingly distinct learning
trajectories across the process diagnostics, even when their return trajectories
are comparatively similar. Conflict and forced-idle behavior also vary
substantially across methods and configurations, while throughput and
assignment diversity reveal differences in how effectively additional scale is
converted into useful task progress. These trajectories provide the
training-time context for the final-policy and controlled-scaling analyses
reported in the main paper.

% Across configurations, return curves often appear more similar across methods than the coordination diagnostics. In contrast, conflict rate and per-agent assignment diversity more clearly separate methods by how they coordinate. 

\subsection{Controlled Scaling Significance Tests}
\label{app:scaling_significance}
Tables~\ref{tab:spatial_scaling_significance}-\ref{tab:agent_scaling_significance}
report statistical tests for the controlled STAT scaling comparisons used in
Section~\ref{sec:rq2}. Each comparison varies one problem factor
while holding the others fixed. We report the direction of change for return
and the four core STAT process diagnostics, with Holm correction applied within
each comparison block.

\subsection{Computational Efficiency}
\label{app:results_computational_efficiency}
% \paragraph{Computational Efficiency.}
We report computational efficiency as additional context for interpreting
scalability. Table~\ref{tab:wallclock_ci95_all_methods} reports wall-clock
training time, while Table~\ref{tab:stepspersec_ci95_all_methods} reports
environment timesteps processed per second. These measurements complement the
coordination analysis by distinguishing behavioral scalability from the
computational cost of learning.

Standard centralized DQN's
% is evaluated only in the three smallest STAT
% configurations because its output layer explicitly enumerates the joint action
% space. 
training throughput decreases substantially as the number of joint
actions grows, illustrating the computational limitation of direct
centralized joint-action enumeration. FDQN avoids this failure mode through
its factorized action representation and remains tractable across all STAT
configurations.

Among the methods evaluated across the full benchmark, computational profiles
also differ substantially. The actor-critic implementations,
particularly MAA2C, generally achieve higher training throughput and shorter
wall-clock times than the value-learning methods, while throughput
for all methods tends to decrease as task or agent scale increases. Because
these methods are implemented using different software frameworks, however,
wall-clock and throughput differences should not be interpreted as controlled
comparisons of algorithmic computational complexity. We report them primarily
to document the practical computational cost of the experiments and to show
that the evaluated methods remain tractable at the reported scales.

% These results highlight that computational tractability and coordination quality should be interpreted jointly: scalable execution does not guarantee low conflict, and strong coordination is most useful when supported by a tractable representation.

\begin{table*}[h]
\centering
\caption{Wall-clock training time across STAT configurations, reported in hours as mean $\pm$ 95\% CI over five seeds. Lower is better.}
\label{tab:wallclock_ci95_all_methods}
\begin{adjustbox}{width=\textwidth}
\begin{tabular}{lccccccccc}
\toprule
\textbf{Training Paradigm} & \multicolumn{2}{c}{\textbf{CTCE}} & \multicolumn{5}{c}{\textbf{CTDE}} & \multicolumn{2}{c}{\textbf{DTDE}} \\
\cmidrule(lr){2-3} \cmidrule(lr){4-8} \cmidrule(lr){9-10}
\textbf{Environments/Methods} & \textbf{DQN} & \textbf{FDQN} & \textbf{VDN} & \textbf{QMIX} & \textbf{QTRAN} & \textbf{MAPPO} & \textbf{MAA2C} & \textbf{IQL} & \textbf{IPPO} \\
\midrule
\textbf{3A-6T-5x3} & 1.87 $\pm$ 0.15 & 0.91 $\pm$ 0.02 & 2.93 $\pm$ 1.09 & 2.62 $\pm$ 0.22 & 2.74 $\pm$ 0.64 & 0.64 $\pm$ 0.02 & 0.35 $\pm$ 0.02 & 2.33 $\pm$ 0.04 & 0.61 $\pm$ 0.06 \\
\textbf{3A-6T-10x6} & 1.84 $\pm$ 0.12 & 1.00 $\pm$ 0.09 & 2.70 $\pm$ 0.72 & 2.25 $\pm$ 0.04 & 2.42 $\pm$ 0.42 & 0.63 $\pm$ 0.02 & 0.35 $\pm$ 0.03 & 2.27 $\pm$ 0.12 & 0.66 $\pm$ 0.04 \\
\textbf{3A-12T-10x6} & 3.45 $\pm$ 0.03 & 0.86 $\pm$ 0.03 & 2.10 $\pm$ 0.20 & 2.07 $\pm$ 0.09 & 2.21 $\pm$ 0.44 & 0.64 $\pm$ 0.04 & 0.34 $\pm$ 0.02 & 2.07 $\pm$ 0.10 & 0.60 $\pm$ 0.01 \\
\textbf{5A-12T-10x6} & -- & 1.09 $\pm$ 0.02 & 2.21 $\pm$ 0.07 & 2.41 $\pm$ 0.11 & 2.31 $\pm$ 0.05 & 0.67 $\pm$ 0.04 & 0.37 $\pm$ 0.03 & 2.54 $\pm$ 0.52 & 0.67 $\pm$ 0.03 \\
\textbf{5A-25T-25x15} & -- & 11.22 $\pm$ 0.46 & 29.42 $\pm$ 0.68 & 30.69 $\pm$ 2.26 & 30.70 $\pm$ 0.97 & 7.93 $\pm$ 0.40 & 4.66 $\pm$ 0.31 & 29.44 $\pm$ 0.57 & 7.94 $\pm$ 0.84 \\
\textbf{5A-25T-50x30} & -- & 11.76 $\pm$ 0.41 & 28.92 $\pm$ 1.48 & 28.91 $\pm$ 0.43 & 29.33 $\pm$ 0.57 & 8.28 $\pm$ 0.26 & 5.01 $\pm$ 0.44 & 28.94 $\pm$ 0.87 & 7.87 $\pm$ 0.29 \\
\textbf{5A-50T-50x30} & -- & 13.68 $\pm$ 0.13 & 32.25 $\pm$ 0.48 & 32.54 $\pm$ 0.40 & 32.80 $\pm$ 0.49 & 9.73 $\pm$ 0.55 & 6.61 $\pm$ 0.78 & 32.20 $\pm$ 0.49 & 10.59 $\pm$ 0.73 \\
\textbf{5A-100T-50x30} & -- & 17.77 $\pm$ 0.11 & 41.73 $\pm$ 0.70 & 41.91 $\pm$ 0.53 & 42.03 $\pm$ 0.69 & 14.27 $\pm$ 0.87 & 11.19 $\pm$ 0.87 & 41.43 $\pm$ 0.70 & 14.12 $\pm$ 1.93 \\
\textbf{9A-25T-50x30} & -- & 16.01 $\pm$ 0.23 & 33.74 $\pm$ 2.85 & 33.69 $\pm$ 0.96 & 33.38 $\pm$ 0.71 & 8.67 $\pm$ 0.63 & 6.10 $\pm$ 0.81 & 33.13 $\pm$ 0.70 & 8.89 $\pm$ 0.52 \\
\bottomrule
\end{tabular}
\end{adjustbox}
\end{table*}

\begin{table*}[h]
\centering
\caption{Training throughput across STAT configurations, reported as environment timesteps per second and averaged over five seeds with 95\% CI. Higher is better.}
\label{tab:stepspersec_ci95_all_methods}
\begin{adjustbox}{width=\textwidth}
\begin{tabular}{lccccccccc}
\toprule
\textbf{Training Paradigm} & \multicolumn{2}{c}{\textbf{CTCE}} & \multicolumn{5}{c}{\textbf{CTDE}} & \multicolumn{2}{c}{\textbf{DTDE}} \\
\cmidrule(lr){2-3} \cmidrule(lr){4-8} \cmidrule(lr){9-10}
\textbf{Environments/Methods} & \textbf{DQN} & \textbf{FDQN} & \textbf{VDN} & \textbf{QMIX} & \textbf{QTRAN} & \textbf{MAPPO} & \textbf{MAA2C} & \textbf{IQL} & \textbf{IPPO} \\
\midrule
\textbf{3A-6T-5x3} & 304.95 $\pm$ 8.4 & 651.25 $\pm$ 1.8 & 585.28 $\pm$ 12.8 & 588.79 $\pm$ 6.4 & 591.51 $\pm$ 10.3 & 864.88 $\pm$ 27.1 & 1572.54 $\pm$ 82.3 & 581.50 $\pm$ 15.0 & 911.89 $\pm$ 91.0 \\
\textbf{3A-6T-10x6} & 326.63 $\pm$ 3.1 & 698.67 $\pm$ 62.0 & 594.15 $\pm$ 23.6 & 603.12 $\pm$ 5.7 & 599.77 $\pm$ 8.9 & 879.79 $\pm$ 29.9 & 1572.55 $\pm$ 136.6 & 584.02 $\pm$ 26.4 & 842.96 $\pm$ 57.4 \\
\textbf{3A-12T-10x6} & 164.04 $\pm$ 1.9 & 708.92 $\pm$ 17.9 & 595.63 $\pm$ 30.4 & 605.16 $\pm$ 6.3 & 592.36 $\pm$ 22.3 & 866.52 $\pm$ 49.7 & 1615.03 $\pm$ 92.6 & 590.13 $\pm$ 26.3 & 926.16 $\pm$ 20.9 \\
\textbf{5A-12T-10x6} & -- & 556.31 $\pm$ 20.3 & 575.19 $\pm$ 7.8 & 573.94 $\pm$ 4.0 & 569.53 $\pm$ 9.8 & 827.83 $\pm$ 51.6 & 1488.55 $\pm$ 127.8 & 566.40 $\pm$ 23.3 & 833.04 $\pm$ 39.5 \\
\textbf{5A-25T-25x15} & -- & 526.52 $\pm$ 22.1 & 548.01 $\pm$ 4.3 & 540.53 $\pm$ 11.4 & 539.69 $\pm$ 6.8 & 701.38 $\pm$ 35.3 & 1195.72 $\pm$ 80.9 & 545.24 $\pm$ 5.3 & 703.86 $\pm$ 75.4 \\
\textbf{5A-25T-50x30} & -- & 515.07 $\pm$ 18.5 & 541.62 $\pm$ 14.6 & 538.73 $\pm$ 7.6 & 539.68 $\pm$ 9.3 & 671.59 $\pm$ 21.4 & 1114.03 $\pm$ 96.3 & 539.20 $\pm$ 8.1 & 706.24 $\pm$ 26.5 \\
\textbf{5A-50T-50x30} & -- & 473.66 $\pm$ 6.8 & 472.20 $\pm$ 7.0 & 469.93 $\pm$ 2.1 & 472.72 $\pm$ 6.6 & 572.06 $\pm$ 32.5 & 846.30 $\pm$ 91.8 & 471.41 $\pm$ 7.5 & 525.82 $\pm$ 35.3 \\
\textbf{5A-100T-50x30} & -- & 421.39 $\pm$ 2.8 & 375.49 $\pm$ 7.0 & 375.01 $\pm$ 5.7 & 377.34 $\pm$ 7.0 & 390.07 $\pm$ 23.2 & 498.18 $\pm$ 38.2 & 376.93 $\pm$ 6.3 & 396.96 $\pm$ 48.1 \\
\textbf{9A-25T-50x30} & -- & 369.71 $\pm$ 5.7 & 486.53 $\pm$ 11.0 & 484.76 $\pm$ 5.1 & 487.88 $\pm$ 3.0 & 642.39 $\pm$ 46.5 & 919.82 $\pm$ 124.6 & 485.53 $\pm$ 6.4 & 625.81 $\pm$ 37.6 \\
\bottomrule
\end{tabular}
\end{adjustbox}
\end{table*}

\subsection{Additional Method-Level Observations}
\label{app:method_level_observations}
We summarize
additional method-level patterns observed across the full set of STAT
configurations. These observations provide context for the main results but
should be interpreted descriptively rather than as causal effects of algorithm
architecture.

\paragraph{Centralized Training and Centralized Execution Methods.}
The centralized methods illustrate the trade-off between joint-action reasoning
and computational tractability. Standard DQN performs competitively in the
smallest configurations, where explicit centralized joint-action reasoning
remains feasible, but becomes intractable at larger scales because its output
layer enumerates the full joint action space. FDQN avoids this failure mode
through a factorized centralized representation and remains tractable across
all STAT configurations. Across the benchmark, FDQN often exhibits relatively
low assignment interference while maintaining competitive task performance,
showing that structured centralized reasoning can support effective
coordination without requiring explicit enumeration of all joint actions.

\paragraph{Centralized Training and Decentralized Execution Methods.}
The CTDE methods span both value-decomposition approaches (VDN, QMIX, and
QTRAN) and centralized-critic actor--critic methods (MAPPO and MAA2C). These
methods remain tractable across the evaluated STAT configurations but exhibit
distinct process profiles. In particular, MAPPO and MAA2C often achieve among
the strongest returns, throughput, and assignment diversity, while not always
minimizing conflicts or forced idle. The value-decomposition methods show
different trade-offs across configurations, with comparable task performance
sometimes arising from substantially different levels of assignment
interference. These results illustrate that strong aggregate performance does
not imply uniformly favorable coordination behavior across all process
dimensions.

\paragraph{Decentralized Training and Decentralized Execution Methods.}
IQL and IPPO provide independent-learning baselines without centralized
training information. IQL frequently exhibits comparatively high conflict and
forced-idle levels, particularly in settings with stronger assignment
interdependence, even when its return remains competitive. IPPO, by contrast,
often achieves strong task performance in the evaluated STAT configurations,
but its process behavior varies across settings. Together, these results show
that decentralized learning does not produce a single coordination pattern and
that outcome performance alone is insufficient to characterize how independent
learners coordinate.

Overall, the method-level results reinforce the central motivation for
process-aware evaluation. No training paradigm uniformly dominates across
return, interference, task completion, participation, and coordination
efficiency. Centralized, CTDE, and decentralized approaches instead exhibit
different behavioral trade-offs across configurations, motivating evaluation
of both task outcomes and the coordination processes through which those
outcomes are achieved.

% \section{Additional Results in LBF and RWARE}
% \label{app:lbf_rware}
\section{Full Cross-Environment Results}
\label{app:lbf_rware}

\subsection{Level-Based Foraging Results}

\begin{table*}[t]
\centering
\small
\setlength{\tabcolsep}{3pt}
\renewcommand{\arraystretch}{1.12}
\caption{Full Level-Based Foraging results. Values show mean performance across
five seeds with 95\% confidence intervals. Arrows indicate the generally
preferred direction for each metric. Failed-load and synchronization-failure
rates characterize unsuccessful coordination during loading, collection
throughput measures successful task completion over time, and participation
balance and partner diversity characterize how successful cooperation is
distributed across agents when applicable. Bold
indicates the best mean within each environment setting.}
\label{tab:lbf_full_results}

\begin{adjustbox}{width=\textwidth}
\begin{tabular}{llcccccccc}
\toprule
\textbf{Setting}
& \textbf{Method}
& \textbf{Return} $\uparrow$
& \textbf{Foods} $\uparrow$
& \textbf{Throughput} $\uparrow$
& \textbf{Failed Load} $\downarrow$
& \textbf{Sync. Failure} $\downarrow$
& \textbf{Sync. Delay} $\downarrow$
& \textbf{Participation} $\uparrow$
& \textbf{Partner Diversity} $\uparrow$ \\
&
&
&
\textbf{Collected}
&
&
\textbf{Rate}
&
\textbf{Rate}
&
&
\textbf{Balance}
&
\textbf{Fraction} \\
\midrule

\multirow{4}{*}{2-agent Full}
& MAPPO
& $0.984 \pm 0.013$
& $1.97 \pm 0.03$
& $0.167 \pm 0.006$
& $0.532 \pm 0.058$
& $0.034 \pm 0.027$
& $1.203 \pm 0.099$
& -- & -- \\

& IPPO
& $0.908 \pm 0.231$
& $1.82 \pm 0.46$
& $0.139 \pm 0.076$
& $0.623 \pm 0.206$
& $0.042 \pm 0.007$
& $1.299 \pm 0.381$
& -- & -- \\

& QMIX
& $0.442 \pm 0.242$
& $0.88 \pm 0.48$
& $0.036 \pm 0.061$
& $0.386 \pm 0.061$
& $0.052 \pm 0.063$
& $\mathbf{0.752 \pm 0.134}$
& -- & -- \\

& IQL
& $\mathbf{0.994 \pm 0.011}$
& $\mathbf{1.99 \pm 0.02}$
& $\mathbf{0.194 \pm 0.011}$
& $\mathbf{0.188 \pm 0.055}$
& $\mathbf{0.004 \pm 0.009}$
& $1.027 \pm 0.033$
& -- & -- \\
\midrule

\multirow{4}{*}{2-agent Partial}
& MAPPO
& $0.981 \pm 0.018$
& $1.96 \pm 0.04$
& $0.120 \pm 0.009$
& $0.505 \pm 0.036$
& $0.060 \pm 0.036$
& $1.352 \pm 0.193$
& -- & -- \\

& IPPO
& $0.981 \pm 0.012$
& $1.96 \pm 0.02$
& $0.122 \pm 0.008$
& $0.507 \pm 0.033$
& $0.057 \pm 0.023$
& $1.385 \pm 0.202$
& -- & -- \\

& QMIX
& $0.593 \pm 0.162$
& $1.19 \pm 0.32$
& $0.075 \pm 0.023$
& $0.330 \pm 0.066$
& $0.044 \pm 0.038$
& $\mathbf{0.852 \pm 0.309}$
& -- & -- \\

& IQL
& $\mathbf{0.990 \pm 0.012}$
& $\mathbf{1.98 \pm 0.02}$
& $\mathbf{0.146 \pm 0.009}$
& $\mathbf{0.242 \pm 0.057}$
& $\mathbf{0.009 \pm 0.008}$
& $1.032 \pm 0.023$
& -- & -- \\
\midrule

\multirow{4}{*}{3-agent Full}
& MAPPO
& $0.984 \pm 0.013$
& $2.96 \pm 0.03$
& $0.160 \pm 0.006$
& $0.563 \pm 0.053$
& $0.016 \pm 0.012$
& $2.443 \pm 0.351$
& $0.932 \pm 0.012$
& $\mathbf{0.853 \pm 0.062}$ \\

& IPPO
& $\mathbf{0.992 \pm 0.009}$
& $\mathbf{2.97 \pm 0.04}$
& $\mathbf{0.171 \pm 0.011}$
& $0.571 \pm 0.029$
& $0.020 \pm 0.011$
& $2.286 \pm 0.405$
& $\mathbf{0.937 \pm 0.021}$
& $\mathbf{0.853 \pm 0.043}$ \\

& QMIX
& $0.033 \pm 0.015$
& $0.16 \pm 0.06$
& $0.003 \pm 0.001$
& $\mathbf{0.206 \pm 0.071}$
& $\mathbf{0.010 \pm 0.010}$
& $\mathbf{0.327 \pm 0.094}$
& $0.006 \pm 0.006$
& $0.003 \pm 0.002$ \\

& IQL
& $0.303 \pm 0.093$
& $1.08 \pm 0.30$
& $0.033 \pm 0.012$
& $0.580 \pm 0.156$
& $0.022 \pm 0.019$
& $1.892 \pm 1.097$
& $0.287 \pm 0.111$
& $0.167 \pm 0.107$ \\
\midrule

\multirow{4}{*}{3-agent Partial}
& MAPPO
& $\mathbf{0.872 \pm 0.047}$
& $\mathbf{2.66 \pm 0.13}$
& $\mathbf{0.123 \pm 0.015}$
& $0.624 \pm 0.096$
& $\mathbf{0.017 \pm 0.014}$
& $1.394 \pm 0.202$
& $\mathbf{0.854 \pm 0.047}$
& $\mathbf{0.731 \pm 0.054}$ \\

& IPPO
& $0.785 \pm 0.068$
& $2.43 \pm 0.16$
& $0.102 \pm 0.015$
& $0.694 \pm 0.060$
& $0.023 \pm 0.013$
& $1.589 \pm 0.334$
& $0.781 \pm 0.058$
& $0.660 \pm 0.055$ \\

& QMIX
& $0.239 \pm 0.045$
& $0.92 \pm 0.15$
& $0.031 \pm 0.011$
& $\mathbf{0.267 \pm 0.048}$
& $0.033 \pm 0.028$
& $\mathbf{0.856 \pm 0.051}$
& $0.226 \pm 0.070$
& $0.090 \pm 0.025$ \\

& IQL
& $0.625 \pm 0.055$
& $2.01 \pm 0.14$
& $0.070 \pm 0.004$
& $0.520 \pm 0.045$
& $0.043 \pm 0.028$
& $1.948 \pm 0.396$
& $0.663 \pm 0.055$
& $0.467 \pm 0.080$ \\
\bottomrule
\end{tabular}
\end{adjustbox}
\end{table*}

Table~\ref{tab:lbf_full_results} reports the complete LBF results across methods and configurations. The clearest outcome-similar comparisons occur in
the two-agent settings. Under full observability, MAPPO and IQL achieve nearly
identical return (0.984 and 0.994), yet IQL exhibits substantially lower
failed-load rate (0.188 vs.\ 0.532), lower synchronization-failure rate
(0.004 vs.\ 0.034), and higher collection throughput (0.194 vs.\ 0.167).
Under partial observability, MAPPO, IPPO, and IQL again achieve similar return,
while IQL shows lower failed-load and synchronization-failure rates together
with higher throughput. These comparisons reinforce that similar task outcomes
can arise from different coordination processes.

The three-agent settings provide complementary evidence. MAPPO and IPPO achieve
similarly strong performance under full observability and also exhibit similar
participation balance and partner diversity, whereas under partial observability
MAPPO achieves higher return, throughput, participation balance, and partner
diversity than IPPO. At the same time, QMIX attains comparatively low failed-load
rates and synchronization delays in some settings despite poor task completion,
showing that favorable values for an individual process metric do not by
themselves imply effective coordination.

\subsection{Robotic Warehouse Results}

\begin{table*}[t]
\centering
\small
\setlength{\tabcolsep}{3pt}
\renewcommand{\arraystretch}{1.12}
\caption{Full Robotic Warehouse results for the 40M-step MAPPO and IPPO runs.
Values show mean performance across five seeds with 95\% confidence intervals.
Delivery events are inferred from positive per-agent rewards before common-reward
aggregation; canceled-movement rate measures blocked forward requests; NO-OP
rate measures inactivity; and delivery contribution CV measures imbalance in
per-agent delivery contributions. Bold indicates the best mean within each
environment setting, except for steps per delivery, which is interpreted
jointly with delivery count because zero-delivery episodes receive a value of
zero under the logging convention.}
\label{tab:rware_full_results}

\begin{adjustbox}{width=\textwidth}
\begin{tabular}{llccccccc}
\toprule
\textbf{Environment}
& \textbf{Method}
& \textbf{Return} $\uparrow$
& \textbf{Delivery} $\uparrow$
& \textbf{Delivery} $\uparrow$
& \textbf{Canceled-Move} $\downarrow$
& \textbf{NO-OP} $\downarrow$
& \textbf{Steps/Delivery} $\downarrow$
& \textbf{Contribution CV} $\downarrow$ \\
&
&
&
\textbf{Events}
&
\textbf{Throughput}
&
\textbf{Rate}
&
\textbf{Rate}
&
&
\\
\midrule

\multirow{2}{*}{\texttt{tiny-2ag}}
& MAPPO
& $\mathbf{3.40 \pm 5.12}$
& $\mathbf{3.40 \pm 5.12}$
& $\mathbf{0.0068 \pm 0.0102}$
& $\mathbf{0.0100 \pm 0.0067}$
& $0.0232 \pm 0.0577$
& $231.8 \pm 132.6$
& $0.244 \pm 0.178$ \\

& IPPO
& $2.72 \pm 1.71$
& $2.72 \pm 1.71$
& $0.0054 \pm 0.0034$
& $0.0142 \pm 0.0242$
& $\mathbf{0.0001 \pm 0.0001}$
& $232.6 \pm 80.9$
& $\mathbf{0.186 \pm 0.121}$ \\
\midrule

\multirow{2}{*}{\texttt{tiny-4ag}}
& MAPPO
& $4.70 \pm 6.22$
& $4.70 \pm 6.22$
& $0.0094 \pm 0.0124$
& $\mathbf{0.0208 \pm 0.0122}$
& $0.0672 \pm 0.1436$
& $49.6 \pm 62.7$
& $\mathbf{0.154 \pm 0.212}$ \\

& IPPO
& $\mathbf{18.53 \pm 12.70}$
& $\mathbf{18.53 \pm 12.70}$
& $\mathbf{0.0371 \pm 0.0254}$
& $0.0290 \pm 0.0051$
& $\mathbf{0.0000 \pm 0.0000}$
& $51.9 \pm 55.9$
& $0.285 \pm 0.203$ \\
\midrule

\multirow{2}{*}{\texttt{small-4ag}}
& MAPPO
& $0.12 \pm 0.31$
& $0.12 \pm 0.31$
& $0.0002 \pm 0.0006$
& $\mathbf{0.0016 \pm 0.0012}$
& $0.1899 \pm 0.5252$
& $42.3 \pm 103.7$
& $\mathbf{0.150 \pm 0.370}$ \\

& IPPO
& $\mathbf{2.71 \pm 1.89}$
& $\mathbf{2.71 \pm 1.89}$
& $\mathbf{0.0054 \pm 0.0038}$
& $0.0227 \pm 0.0225$
& $\mathbf{0.0000 \pm 0.0000}$
& $131.9 \pm 93.6$
& $0.280 \pm 0.207$ \\
\bottomrule
\end{tabular}
\end{adjustbox}
\end{table*}

Table~\ref{tab:rware_full_results} reports the complete RWARE results for the
40M-step MAPPO and IPPO runs. The process diagnostics reveal a different type
of coordination tradeoff than in LBF. In \texttt{small-4ag}, MAPPO exhibits a
much lower canceled-movement rate than IPPO (0.0016 vs.\ 0.0227), but also
almost no delivery activity, with only 0.12 delivery events and near-zero
throughput. IPPO experiences more blocked forward requests while completing
substantially more delivery events and achieving much higher throughput.

A similar pattern appears in \texttt{tiny-4ag} where IPPO achieves substantially
higher return and delivery throughput despite a higher canceled-movement rate
than MAPPO. These results show that low movement interference can reflect
efficient coordination, but can also reflect limited engagement with the task.
The NO-OP and contribution diagnostics provide additional context for
distinguishing these cases, supporting the broader framework principle that
process metrics should be interpreted jointly with task completion rather than
in isolation.

\begin{table*}[t]
\centering
\footnotesize
\setlength{\tabcolsep}{4pt}
\renewcommand{\arraystretch}{1.08}
\caption{Statistical tests for controlled spatial-extent scaling in STAT.
Entries show the direction of change from the first configuration to the
second. $\uparrow$ and $\downarrow$ indicate increases and decreases,
respectively. $^{*}$ denotes Holm-corrected $p<0.05$ under Welch's two-sample
$t$-test within each comparison block, and superscript $\mathrm{ns}$ denotes a
non-significant change.}
\label{tab:spatial_scaling_significance}

\begin{tabular}{llccccc}
\toprule
\textbf{Comparison}
& \textbf{Method}
& \textbf{Return}
& \textbf{Conflicts/Opp.}
& \textbf{Throughput}
& \textbf{Diversity/Agent}
& \textbf{Forced Idle/Opp.} \\
\midrule

\multirow{9}{*}{Baseline}
& DQN   & $\downarrow^{*}$ & $\downarrow^{\mathrm{ns}}$ & $\downarrow^{*}$ & $\downarrow^{*}$ & $\downarrow^{\mathrm{ns}}$ \\
& FDQN  & $\downarrow^{*}$ & $\uparrow^{\mathrm{ns}}$   & $\downarrow^{*}$ & $\downarrow^{*}$ & $\uparrow^{\mathrm{ns}}$ \\
& VDN   & $\downarrow^{*}$ & $\uparrow^{\mathrm{ns}}$   & $\downarrow^{*}$ & $\downarrow^{*}$ & $\uparrow^{\mathrm{ns}}$ \\
& QMIX  & $\downarrow^{*}$ & $\uparrow^{\mathrm{ns}}$   & $\downarrow^{*}$ & $\downarrow^{*}$ & $\uparrow^{\mathrm{ns}}$ \\
& QTRAN & $\downarrow^{*}$ & $\downarrow^{\mathrm{ns}}$ & $\downarrow^{*}$ & $\downarrow^{*}$ & $\downarrow^{\mathrm{ns}}$ \\
& MAPPO & $\downarrow^{*}$ & $\uparrow^{\mathrm{ns}}$   & $\downarrow^{*}$ & $\downarrow^{*}$ & $\uparrow^{\mathrm{ns}}$ \\
& MAA2C & $\downarrow^{*}$ & $\downarrow^{\mathrm{ns}}$ & $\downarrow^{*}$ & $\downarrow^{*}$ & $\downarrow^{\mathrm{ns}}$ \\
& IQL   & $\downarrow^{*}$ & $\uparrow^{\mathrm{ns}}$   & $\downarrow^{*}$ & $\downarrow^{*}$ & $\uparrow^{\mathrm{ns}}$ \\
& IPPO  & $\downarrow^{*}$ & $\uparrow^{\mathrm{ns}}$   & $\downarrow^{*}$ & $\downarrow^{*}$ & $\uparrow^{\mathrm{ns}}$ \\
\midrule

\multirow{9}{*}{Extreme}
& DQN   & -- & -- & -- & -- & -- \\
& FDQN  & $\downarrow^{*}$ & $\downarrow^{*}$ & $\downarrow^{*}$ & $\downarrow^{*}$ & $\downarrow^{*}$ \\
& VDN   & $\downarrow^{*}$ & $\downarrow^{*}$ & $\downarrow^{*}$ & $\downarrow^{*}$ & $\downarrow^{*}$ \\
& QMIX  & $\downarrow^{*}$ & $\downarrow^{*}$ & $\downarrow^{*}$ & $\downarrow^{*}$ & $\downarrow^{\mathrm{ns}}$ \\
& QTRAN & $\downarrow^{*}$ & $\downarrow^{*}$ & $\downarrow^{*}$ & $\downarrow^{*}$ & $\downarrow^{*}$ \\
& MAPPO & $\downarrow^{*}$ & $\downarrow^{*}$ & $\downarrow^{*}$ & $\downarrow^{*}$ & $\downarrow^{*}$ \\
& MAA2C & $\downarrow^{*}$ & $\downarrow^{*}$ & $\downarrow^{*}$ & $\downarrow^{*}$ & $\downarrow^{*}$ \\
& IQL   & $\downarrow^{*}$ & $\downarrow^{*}$ & $\downarrow^{*}$ & $\downarrow^{*}$ & $\downarrow^{*}$ \\
& IPPO  & $\downarrow^{*}$ & $\downarrow^{*}$ & $\downarrow^{*}$ & $\downarrow^{*}$ & $\downarrow^{*}$ \\
\bottomrule
\end{tabular}

\vspace{0.4em}
\raggedright
\footnotesize
Baseline compares 3A--6T--5x3 $\rightarrow$ 3A--6T--10x6.
Extreme compares 5A--25T--25x15 $\rightarrow$ 5A--25T--50x30.
\end{table*}

% ============================================================
% Task-count scaling
% ============================================================

\begin{table*}[t]
\centering
\footnotesize
\setlength{\tabcolsep}{4pt}
\renewcommand{\arraystretch}{1.08}
\caption{Statistical tests for controlled task-count scaling in STAT.
Entries show the direction of change from the first configuration to the
second. $\uparrow$ and $\downarrow$ indicate increases and decreases,
respectively. $^{*}$ denotes Holm-corrected $p<0.05$ under Welch's two-sample
$t$-test within each comparison block, and superscript $\mathrm{ns}$ denotes a
non-significant change.}
\label{tab:task_scaling_significance}

\begin{tabular}{llccccc}
\toprule
\textbf{Comparison}
& \textbf{Method}
& \textbf{Return}
& \textbf{Conflicts/Opp.}
& \textbf{Throughput}
& \textbf{Diversity/Agent}
& \textbf{Forced Idle/Opp.} \\
\midrule

\multirow{9}{*}{Baseline}
& DQN   & $\uparrow^{*}$ & $\uparrow^{\mathrm{ns}}$   & $\uparrow^{*}$ & $\uparrow^{*}$             & $\uparrow^{\mathrm{ns}}$ \\
& FDQN  & $\uparrow^{*}$ & $\downarrow^{\mathrm{ns}}$ & $\uparrow^{*}$ & $\uparrow^{*}$             & $\downarrow^{\mathrm{ns}}$ \\
& VDN   & $\uparrow^{*}$ & $\uparrow^{\mathrm{ns}}$   & $\uparrow^{*}$ & $\uparrow^{*}$             & $\uparrow^{\mathrm{ns}}$ \\
& QMIX  & $\uparrow^{*}$ & $\uparrow^{\mathrm{ns}}$   & $\uparrow^{*}$ & $\uparrow^{\mathrm{ns}}$   & $\uparrow^{\mathrm{ns}}$ \\
& QTRAN & $\uparrow^{*}$ & $\downarrow^{\mathrm{ns}}$ & $\uparrow^{*}$ & $\uparrow^{*}$             & $\uparrow^{\mathrm{ns}}$ \\
& MAPPO & $\uparrow^{*}$ & $\uparrow^{\mathrm{ns}}$   & $\uparrow^{*}$ & $\uparrow^{*}$             & $\uparrow^{\mathrm{ns}}$ \\
& MAA2C & $\uparrow^{*}$ & $\downarrow^{*}$           & $\uparrow^{*}$ & $\uparrow^{*}$             & $\downarrow^{*}$ \\
& IQL   & $\uparrow^{*}$ & $\downarrow^{*}$           & $\uparrow^{*}$ & $\uparrow^{*}$             & $\downarrow^{*}$ \\
& IPPO  & $\uparrow^{*}$ & $\uparrow^{\mathrm{ns}}$   & $\uparrow^{*}$ & $\uparrow^{*}$             & $\uparrow^{\mathrm{ns}}$ \\
\midrule

\multirow{9}{*}{Extreme 25$\rightarrow$50}
& DQN   & -- & -- & -- & -- & -- \\
& FDQN  & $\uparrow^{*}$ & $\downarrow^{\mathrm{ns}}$ & $\uparrow^{*}$ & $\uparrow^{*}$ & $\downarrow^{\mathrm{ns}}$ \\
& VDN   & $\uparrow^{*}$ & $\downarrow^{\mathrm{ns}}$ & $\uparrow^{*}$ & $\uparrow^{*}$ & $\downarrow^{\mathrm{ns}}$ \\
& QMIX  & $\uparrow^{*}$ & $\downarrow^{*}$           & $\uparrow^{*}$ & $\uparrow^{*}$ & $\downarrow^{\mathrm{ns}}$ \\
& QTRAN & $\uparrow^{*}$ & $\downarrow^{*}$           & $\uparrow^{*}$ & $\uparrow^{*}$ & $\downarrow^{*}$ \\
& MAPPO & $\uparrow^{*}$ & $\downarrow^{*}$           & $\uparrow^{*}$ & $\uparrow^{*}$ & $\downarrow^{*}$ \\
& MAA2C & $\uparrow^{*}$ & $\downarrow^{*}$           & $\uparrow^{*}$ & $\uparrow^{*}$ & $\downarrow^{*}$ \\
& IQL   & $\uparrow^{*}$ & $\downarrow^{*}$           & $\uparrow^{*}$ & $\uparrow^{*}$ & $\downarrow^{*}$ \\
& IPPO  & $\uparrow^{*}$ & $\downarrow^{*}$           & $\uparrow^{*}$ & $\uparrow^{*}$ & $\downarrow^{*}$ \\
\midrule

\multirow{9}{*}{Extreme 50$\rightarrow$100}
& DQN   & -- & -- & -- & -- & -- \\
& FDQN  & $\uparrow^{*}$ & $\downarrow^{*}$           & $\uparrow^{*}$ & $\uparrow^{*}$ & $\downarrow^{*}$ \\
& VDN   & $\uparrow^{*}$ & $\downarrow^{\mathrm{ns}}$ & $\uparrow^{*}$ & $\uparrow^{*}$ & $\uparrow^{\mathrm{ns}}$ \\
& QMIX  & $\uparrow^{*}$ & $\downarrow^{\mathrm{ns}}$ & $\uparrow^{*}$ & $\uparrow^{*}$ & $\downarrow^{\mathrm{ns}}$ \\
& QTRAN & $\uparrow^{*}$ & $\downarrow^{*}$           & $\uparrow^{*}$ & $\uparrow^{*}$ & $\downarrow^{*}$ \\
& MAPPO & $\uparrow^{*}$ & $\downarrow^{*}$           & $\uparrow^{*}$ & $\uparrow^{*}$ & $\downarrow^{*}$ \\
& MAA2C & $\uparrow^{*}$ & $\downarrow^{*}$           & $\uparrow^{*}$ & $\uparrow^{*}$ & $\downarrow^{*}$ \\
& IQL   & $\uparrow^{*}$ & $\downarrow^{*}$           & $\uparrow^{*}$ & $\uparrow^{*}$ & $\downarrow^{*}$ \\
& IPPO  & $\uparrow^{*}$ & $\downarrow^{*}$           & $\uparrow^{*}$ & $\uparrow^{*}$ & $\downarrow^{*}$ \\
\bottomrule
\end{tabular}

\vspace{0.4em}
\raggedright
\footnotesize
Baseline compares 3A--6T--10x6 $\rightarrow$ 3A--12T--10x6.
Extreme 25$\rightarrow$50 compares 5A--25T--50x30 $\rightarrow$
5A--50T--50x30.
Extreme 50$\rightarrow$100 compares 5A--50T--50x30 $\rightarrow$
5A--100T--50x30.
\end{table*}

% ============================================================
% Agent-count scaling
% ============================================================

\begin{table*}[t]
\centering
\footnotesize
\setlength{\tabcolsep}{4pt}
\renewcommand{\arraystretch}{1.08}
\caption{Statistical tests for controlled agent-count scaling in STAT.
Entries show the direction of change from the first configuration to the
second. $\uparrow$ and $\downarrow$ indicate increases and decreases,
respectively. $^{*}$ denotes Holm-corrected $p<0.05$ under Welch's two-sample
$t$-test within each comparison block, and superscript $\mathrm{ns}$ denotes a
non-significant change.}
\label{tab:agent_scaling_significance}

\begin{tabular}{llccccc}
\toprule
\textbf{Comparison}
& \textbf{Method}
& \textbf{Return}
& \textbf{Conflicts/Opp.}
& \textbf{Throughput}
& \textbf{Diversity/Agent}
& \textbf{Forced Idle/Opp.} \\
\midrule

\multirow{9}{*}{Baseline}
& DQN   & -- & -- & -- & -- & -- \\
& FDQN  & $\uparrow^{\mathrm{ns}}$   & $\uparrow^{*}$             & $\uparrow^{*}$ & $\downarrow^{*}$ & $\uparrow^{*}$ \\
& VDN   & $\uparrow^{\mathrm{ns}}$   & $\uparrow^{\mathrm{ns}}$   & $\uparrow^{*}$ & $\downarrow^{*}$ & $\uparrow^{*}$ \\
& QMIX  & $\downarrow^{\mathrm{ns}}$ & $\uparrow^{*}$             & $\uparrow^{*}$ & $\downarrow^{*}$ & $\uparrow^{*}$ \\
& QTRAN & $\downarrow^{\mathrm{ns}}$ & $\uparrow^{*}$             & $\uparrow^{*}$ & $\downarrow^{*}$ & $\uparrow^{*}$ \\
& MAPPO & $\downarrow^{*}$           & $\uparrow^{*}$             & $\uparrow^{*}$ & $\downarrow^{*}$ & $\uparrow^{*}$ \\
& MAA2C & $\downarrow^{*}$           & $\uparrow^{*}$             & $\uparrow^{*}$ & $\downarrow^{*}$ & $\uparrow^{*}$ \\
& IQL   & $\downarrow^{*}$           & $\uparrow^{*}$             & $\uparrow^{*}$ & $\downarrow^{*}$ & $\uparrow^{*}$ \\
& IPPO  & $\downarrow^{*}$           & $\uparrow^{*}$             & $\uparrow^{*}$ & $\downarrow^{*}$ & $\uparrow^{*}$ \\
\midrule

\multirow{9}{*}{Extreme}
& DQN   & -- & -- & -- & -- & -- \\
& FDQN  & $\uparrow^{*}$ & $\uparrow^{*}$             & $\uparrow^{*}$ & $\downarrow^{*}$ & $\uparrow^{*}$ \\
& VDN   & $\uparrow^{*}$ & $\uparrow^{*}$             & $\uparrow^{*}$ & $\downarrow^{*}$ & $\uparrow^{*}$ \\
& QMIX  & $\uparrow^{*}$ & $\uparrow^{*}$             & $\uparrow^{*}$ & $\downarrow^{*}$ & $\uparrow^{*}$ \\
& QTRAN & $\uparrow^{*}$ & $\uparrow^{*}$             & $\uparrow^{*}$ & $\downarrow^{*}$ & $\uparrow^{*}$ \\
& MAPPO & $\uparrow^{*}$ & $\uparrow^{*}$             & $\uparrow^{*}$ & $\downarrow^{*}$ & $\uparrow^{*}$ \\
& MAA2C & $\uparrow^{*}$ & $\uparrow^{*}$             & $\uparrow^{*}$ & $\downarrow^{*}$ & $\uparrow^{*}$ \\
& IQL   & $\uparrow^{*}$ & $\uparrow^{\mathrm{ns}}$   & $\uparrow^{*}$ & $\downarrow^{*}$ & $\uparrow^{*}$ \\
& IPPO  & $\uparrow^{*}$ & $\uparrow^{*}$             & $\uparrow^{*}$ & $\downarrow^{*}$ & $\uparrow^{*}$ \\
\bottomrule
\end{tabular}

\vspace{0.4em}
\raggedright
\footnotesize
Baseline compares 3A--12T--10x6 $\rightarrow$ 5A--12T--10x6.
Extreme compares 5A--25T--50x30 $\rightarrow$ 9A--25T--50x30.
\end{table*}

\end{document}